\documentclass[twocolumn]{aastex63}

\usepackage{textcomp, gensymb}
\usepackage{xcolor}
\usepackage{wrapfig}
\usepackage{tabularx}
\usepackage{changepage}
\usepackage{placeins}
\usepackage[figuresright]{rotating}

\let\tablenum\relax
\usepackage{siunitx}
\usepackage{graphicx,bm,times,amsmath,array,float}
\received{\today}

\shorttitle{CO(1--0) and [\ion{C}{1}](1--0) as molecular gas tracers in high--$z$ SMGs}
\shortauthors{Frias Castillo et al.}

\graphicspath{{./}{figures/}}

\begin{document}

\title{A Comparative Study of the Ground State Transitions of CO and [\ion{C}{1}] as Molecular Gas Tracers at High Redshift}

\correspondingauthor{Marta Frias Castillo}
\email{friascastillom@strw.leidenuniv.nl}

\author[0000-0002-9278-7028]{Marta Frias Castillo}
\affiliation{Leiden Observatory, Leiden University, P.O. Box 9513, 2300 RA Leiden, The Netherlands}

\author[0000-0002-1383-0746]{Matus Rybak}
\affiliation{Faculty of Electrical Engineering, Mathematics and Computer Science,Delft University of Technology, Mekelweg 4, 2628 CD Delft, the Netherlands}
\affiliation{Leiden Observatory, Leiden University, P.O. Box 9513, 2300 RA Leiden, The Netherlands}
\affiliation{SRON - Netherlands Institute for Space Research, Niels Bohrweg 4, 2333 CA Leiden, The Netherlands}

\author[0000-0001-6586-8845]{Jacqueline A. Hodge}
\affiliation{Leiden Observatory, Leiden University, P.O. Box 9513, 2300 RA Leiden, The Netherlands}

\author[0000-0001-5434-5942]{Paul van der Werf}
\affiliation{Leiden Observatory, Leiden University, P.O. Box 9513, 2300 RA Leiden, The Netherlands}

\author[0000-0003-3037-257X]{Ian Smail}
\affiliation{Centre for Extragalactic Astronomy, Department of Physics, Durham University, South Road, Durham DH1 3LE, UK}

\author[0000-0002-5353-1775]{Joshua Butterworth}
\affiliation{Leiden Observatory, Leiden University, P.O. Box 9513, 2300 RA Leiden, The Netherlands}

\author{Jasper Jansen}
\affiliation{Leiden Observatory, Leiden University, P.O. Box 9513, 2300 RA Leiden, The Netherlands}

\author[0000-0001-9143-1495]{Theodoros Topkaras}
\affiliation{ I. Physikalisches Institut, Universit\"at zu K\"oln, Z\"ulpicher Str. 77, 50937 K\"oln, Germany}

\author[0000-0002-3805-0789]{Chian-Chou Chen}
\affiliation{Academia Sinica Institute of Astronomy and Astrophysics (ASIAA), No.1, Section 4, Roosevelt Road, Taipei 10617, Taiwan}

\author{Scott C. Chapman}
\affiliation{Department of Physics and Atmospheric Science, Dalhousie University, Halifax, Halifax, NS B3H 3J5, Canada}

\author[0000-0003-4678-3939]{Axel Weiss}
\affiliation{Max Planck Institut f\"ur Radioastronomie, Auf dem H\"ugel 69, D-53121 Bonn, Germany}

\author[0000-0002-4205-9567]{Hiddo Algera}
\affiliation{Hiroshima Astrophysical Science Center, Hiroshima University, 1-3-1 Kagamiyama, Higashi-Hiroshima, Hiroshima 739-8526, Japan}
\affiliation{National Astronomical Observatory of Japan, 2-21-1, Osawa, Mitaka, Tokyo, Japan}

\author[0000-0002-3272-7568]{Jack E. Birkin}
\affiliation{Department of Physics and Astronomy, Texas A\&M University, 4242 TAMU, College Station, TX 77843-4242, USA}
\affiliation{George P. and Cynthia Woods Mitchell Institute for Fundamental Physics and Astronomy, Texas A\&M University, 4242 TAMU, College Station, TX 77843-4242, USA}

\author[0000-0001-9759-4797]{Elisabete da Cunha}
\affiliation{International Centre for Radio Astronomy Research, University of Western Australia, 35 Stirling Hwy, Crawley, WA 6009, Australia}
\affiliation{ARC Centre of Excellence for All Sky Astrophysics in 3 Dimensions (ASTRO 3D)}

\author[0000-0003-3921-3313]{Jianhang Chen}
\affiliation{European Southern Observatory (ESO), Karl-Schwarzschild-Stra\ss e 2, 85748 Garching bei M\"unchen, Germany }

\author[0000-0001-7147-3575]{Helmut Dannerbauer}
\affiliation{Instituto Astrof\'isica de Canarias (IAC), E-38205 La Laguna, Tenerife, Spain}
\affiliation{Dpto. Astrof\'isica, Universidad de la Laguna, E-38206 La Laguna, Tenerife, Spain}

\author[0000-0002-2640-5917]{E.F. Jim\'enez-Andrade}
\affiliation{Instituto de Radioastronom\'ia y Astrof\'isica, Universidad Nacional Aut\'onoma de M\'exico, Antigua Carretera a P\'atzcuaro \# 8701,\\ Ex-Hda. San Jos\'e de la Huerta, Morelia, Michoac\'an, M\'exico C.P. 58089}

\author{Soh Ikarashi}
\affiliation{Department of Physics, General Studies, College of Engineering, Nihon University, 1 Nakagawara, Tokusada, Tamuramachi, Koriyama, Fukushima, 963-8642, Japan}
\affiliation{National Astronomical Observatory of Japan, 2-21-1 Osawa, Mitaka, Tokyo, 181-8588, Japan}

\author[0000-0002-5247-6639]{Cheng-Lin Liao}
\affiliation{Academia Sinica Institute of Astronomy and Astrophysics (ASIAA), No.1, Section 4, Roosevelt Road, Taipei 10617, Taiwan}
\affiliation{Graduate Institute of Astrophysics, National Taiwan University, Taipei 10617, Taiwan}

\author[0000-0001-7089-7325]{Eric J.\,Murphy}
\affiliation{National Radio Astronomy Observatory, 520 Edgemont Road, Charlottesville, VA 22903, USA}

\author[0000-0003-1192-5837]{A.M. Swinbank}
\affiliation{Centre for Extragalactic Astronomy, Department of Physics, Durham University, South Road, Durham DH1 3LE, UK}
 
\author[0000-0003-4793-7880]{Fabian Walter}
\affiliation{Max Planck Institute for Astronomy, K\"onigstuhl 17, 69117,  Heidelberg, Germany}
\affiliation{National Radio Astronomy Observatory, Pete V. Domenici Array Science Center, P.O. Box O, Socorro, NM 87801, USA}

\author[0000-0003-0085-6346]{Gabriela Calistro Rivera}
\affiliation{European Southern Observatory (ESO), Karl-Schwarzschild-Stra\ss e 2, 85748 Garching bei M\"unchen, Germany }

\author[0000-0001-5118-1313]{R.\,J.~Ivison}
\affiliation{European Southern Observatory (ESO), Karl-Schwarzschild-Stra\ss e 2, D-85748 Garching, Germany}
\affiliation{School of Cosmic Physics, Dublin Institute for Advanced Studies, 31 Fitzwilliam Place, Dublin D02 XF86, Ireland}
\affiliation{Institute for Astronomy, University of Edinburgh, Royal Observatory, Blackford Hill, Edinburgh EH9 3HJ, UK}
\affiliation{ARC Centre of Excellence for All Sky Astrophysics in 3 Dimensions (ASTRO 3D)}

\author[0000-0003-3021-8564]{Claudia del P. Lagos}
\affiliation{International Centre for Radio Astronomy Research (ICRAR), M468, University of Western Australia, 35 Stirling Hwy, Crawley, WA 6009, Australia}
\affiliation{ARC Centre of Excellence for All Sky Astrophysics in 3 Dimensions (ASTRO 3D)}
\affiliation{Cosmic Dawn Center (DAWN), R\aa dmandsgade 62, DK-2200 København, Denmark}

\begin{abstract} 

The CO(1--0) and [\ion{C}{1}](1--0) emission lines are well-established tracers of cold molecular gas mass in local galaxies. At high redshift, where the interstellar medium (ISM) is likely to be denser, there have been limited direct comparisons of both ground state transitions. Here we present a study of CO(1--0) and [\ion{C}{1}](1--0) emission in a sample of 20 unlensed dusty, star-forming galaxies at $z~=~2-5$. The CO(1--0)/[\ion{C}{1}](1--0) ratio is constant up to at least $z=5$, supporting the use of [CI](1-0) as a gas mass tracer. 
PDR modelling of the available data indicates a median H$_2$ density of log$(n~[$cm$^{-3}])=4.7\pm0.2$, and UV radiation field log$(G_{\mathrm{UV}}~[G_0])=3.2\pm0.2$. We use the CO(1--0), [\ion{C}{1}](1--0) and 3mm dust continuum measurements to cross--calibrate the respective gas mass conversion factors, finding no dependence of these factors on either redshift or infrared luminosity. Assuming a variable CO conversion factor then implies [\ion{C}{1}] and dust conversion factors that differ from canonically assumed values but are consistent with the solar/super–solar metallicities expected for our sources.
Radiative transfer modelling shows that the warmer CMB at high redshift can significantly affect the [\ion{C}{1}] as well as CO emission, which can change the derived molecular gas masses by up to 70\% for the coldest kinetic gas temperatures expected. Nevertheless, we show that the magnitude of the effect on the ratio of the tracers is within the known scatter of the $L'_\mathrm{CO}-L'_\mathrm{[CI]}$ relation. Further determining the absolute decrease of individual line intensities will require well--sampled spectral line energy distributions (SLEDs) to model the gas excitation conditions in more detail.

\end{abstract}


\section{Introduction}

The cold gas content of galaxies is expected to be one of the main drivers of the cosmic star formation rate (SFR) density of the Universe. Deep, blind surveys over modest volumes of the cosmic gas density reveal an evolutionary trend resembling that of the cosmic SFR density \citep[e.g.,][]{decarli2020,walter2020}, with both quantities peaking at $z=1-3$. Moreover, the gas fractions of star--forming galaxies have been shown to increase at earlier times (from $\sim$5\% at $z\sim$0 to $\sim$50\% at $z\sim$3), supporting the fact that the increased availability of molecular gas reservoirs at $z=$1--3 is likely to be the primary factor driving the higher SFRs seen in the early Universe \citep[e.g.,][]{tacconi2010,tacconi2020,saintonge2013,bethermin2015,decarli2020,dessauges-zavadsky2020,birkin2020}.

The main component of the molecular gas, molecular hydrogen (H$_2$), cannot be excited in its rotational/vibrational transitions in the low temperatures of the interstellar medium (ISM) of galaxies due to the large separation between its energy levels ($\sim$500 K). Therefore, studies have traditionally relied on observations of carbon monoxide ($^{12}$CO, hereafter CO) emission lines, particularly the ground $J=1-0$ transition, as an alternative tracer of the cold molecular gas reservoirs of galaxies \citep[e.g.,][]{solomon2005,hainline2006,frayer2011,ivison2011,harris2012,thompson2012,bothwell2014,saintonge2017,huynh2017}. However, this tracer requires the assumption of a CO-to-H$_2$ conversion factor, $\alpha_\mathrm{CO}$, which increases with decreasing metallicity \citep[e.g.,][]{bolatto2013} and has been suggested to vary between normal, star--forming galaxies and starbursts \citep{downes1998}. Due to the lack of sufficient data to independently derive $\alpha_\mathrm{CO}$ in most galaxies, it is often assumed to be bimodal, with $\alpha_\mathrm{CO}\sim$3.6 M$_\odot$ (K km s$^{-1}$ pc$^2$)$^{-1}$ for galaxies on the main sequence and $\sim$1 M$_\odot$ (K km s$^{-1}$ pc$^2$)$^{-1}$ for ULIRG--like, starburst galaxies \citep[see][for a review]{bolatto2013}. 

Further, due to the technical challenges of detecting the faint CO(1--0) line emission at high redshift, studies of molecular gas at high redshift commonly rely on the brighter mid--$J_\mathrm{up}$ CO lines \citep[e.g.,][]{bothwell2013,tacconi2013,boogaard2020,decarli2020,birkin2020}. Obtaining molecular gas masses from these lines then depends on line excitation corrections that introduce additional uncertainty \citep[e.g.,][]{bothwell2013,carilli-walter2013,sharon2016,boogaard2020,riechers2020a-vlaspecs,birkin2020,friascastillo2023}. At high redshift ($z>$ 3), this is further complicated by the cosmic microwave background (CMB), which acts as an additional excitation source for the gas and reduces the contrast between the background and the CO line emission. This can bias the information derived from line intensity measurements and imaging of the CO gas distribution of high-redshift galaxies \citep{dacunha2013,zhang2016}.

A separate way to indirectly estimate molecular gas masses is via the long wavelength dust continuum emission \citep[e.g.,][]{hildebrand1983,scoville2016,liu2019,kaasinen2019,wang2022}. Observations of the dust continuum are generally less observationally expensive, allowing for larger samples to be studied. This method, however, requires the assumption
of a mass-weighted cold dust temperature (typically assumed to be $T_\mathrm{dust}$ = 25 K, which is claimed to be a representative value for both local
star-forming galaxies and high-redshift galaxies, \citealt{scoville2016,scoville2017}), and knowledge (or the assumption) of the gas--to--dust abundance ratio, which is dependent on a number of factors, e.g., optical depth, geometry, galaxy metallicity and dust grain properties \citep[e.g.,][]{bolatto2013,popping2022,popping2023}.

The ground state atomic carbon [\ion{C}{1}]($^3$P$_1$–$^3$P$_0$) emission line ($\nu_\mathrm{rest}$=492.161 GHz, hereafter [\ion{C}{1}](1-0)), may provide a promising alternative direct molecular gas tracer \citep[e.g.,][]{papadopoulos2004,papadopoulos2012,walter2011}. Early, simple plane-parallel modeling of star--forming regions predicted that the [\ion{C}{1}] emission arose only from narrow gas layers between the CO and [\ion{C}{2}] emitting regions \citep{tielens1985}. However, subsequent observational work on nearby galaxies \citep[e.g.,][]{keene1997,ojha2001,jiao2019} and theoretical work \citep{tomasetti2014} have shown CO and [\ion{C}{1}] emission to be fully concomitant, with tightly correlated intensities over a wide range of environments. [\ion{C}{1}] emission is typically found to be optically thin \citep{ikeda2002,weiss2003,harrington2021}, which means it can probe high column density environments, while its energy above ground is sufficiently low (24 K) to trace the bulk of the cold gas content, proving itself as a powerful alternative tool to measuring total molecular gas masses. Furthermore, when combined with CO line ratios, the [\ion{C}{1}] lines reflect ISM properties such as density and UV radiation field strength \citep[e.g.][]{hollenbach1999,kaufman2006,alaghband2013,israel2015,bothwell2017,andreani2018,valentino2018,valentino2020}. 

As with CO(1--0), the warmer CMB can have a significant impact at high redshift, but so far this effect has not been studied observationally.
At high redshift, the [\ion{C}{1}] emission line luminosity is correlated with that of CO in a variety of galaxy populations, from star--forming galaxies on the main--sequence \citep[MS,][]{popping2017,bourne2019,valentino2018,valentino2020,dunne2022} to more extreme submillimeter galaxies \citep[SMGs,][]{smail1997,hughes1998,hodge+dacunha2020} and quasar (QSO) hosts at $z\sim2.5$ up to $z\sim4$ \citep{walter2011,alaghband2013,bothwell2017,yang2017,andreani2018,birkin2020,gururajan2023}. Many of these studies, however, suffer from one or more of three shortcomings. Firstly, several use galaxy--scale strong gravitationally lensed systems. This introduces potential differential magnification effects and biases studies towards the most extreme sources \citep[e.g.,][]{yang2017}. Second, in many non--lensed galaxies, CO(1--0) is too faint to detect in reasonable exposure times. As a result, the majority of studies rely on $J_\mathrm{up}>2$ CO line observations and need to assume uncertain excitation corrections, which result in uncertainties in the derived CO(1--0) luminosity of up to 0.5~dex \citep{friascastillo2023}.
Finally, the few studies that have both [\ion{C}{1}](1--0) and CO(1--0) ground-state transitions have relied on individual, targeted sources \citep[e.g.,][]{danielson2011}, and it is thus challenging to draw statistically significant conclusions. Recently, \citet{dunne2022} compiled all of the available literature studies using CO(1--0), [\ion{C}{1}](1--0) and dust  to provide a cross-calibration for metal-rich galaxies from $z\sim0$ to $z\sim5$, arguing that a single [\ion{C}{1}]-to-H$_2$ conversion factor is applicable for both main sequence and starburst galaxies. However, the number of unlensed $z>2$ sources with both ground-state CO and [\ion{C}{1}] observations in their compilation was still $<$10, suggesting the need for further investigation.

\begin{table*}[!htb]
\centering
\caption{Target sample \label{tab:sample} and details of the new [\ion{C}{1}](1--0) ALMA observations.}
\begin{tabular}{@{}lcccccccc @{}}
 \hline \hline
Target & R.A. & Dec & $z$ & Date & rms per channel $^a$ & Beam & Phase calibrator & Flux calibrator \\
& [J2000] & [J2000] & &  & [mJy beam$^{-1}$] & [maj $\times$ min, PA] & &\\
\hline
AS2COS0001.1 & 10:00:08.0 & +02:26:12.3 & 4.625 & 05-03-2022 & 0.32 & 4.6$''\times$3.3$''$,$-$66\degree & J0948+0022 & J1058+0133\\
AS2COS0002.1 & 10:00:15.6 & +02:15:49.0 & 4.595 & 05-03-2022 & 0.33 & 4.7$''\times$3.2$''$,$-$66\degree & J0948+0022 & J1058+0133\\
AS2COS0008.1 & 10:02:49.2 & +02:32:55.5 & 3.581 & 02-03-2022 & 0.25 & 3.5$''\times$3.0$''$,85\degree & J1008-0029 & J1058+0133\\
AS2COS0009.1 & 10:00:28.7 & +02:32:03.6 & 2.260 & 02-03-2022 & 0.25 & 2.42$''\times$2.1$''$,$-$73\degree & J0948+0022 & J1058+0133\\
AS2COS0014.1 & 10:01:41.0 & +02:04:04.9 & 2.921 & 21-03-2022 & 0.39 & 3.7$''\times$2.1$''$,$-$60\degree & J0948+0022 & J1058+0133\\
AS2COS0023.1 & 09:59:42.9 & +02:29:38.2 & 4.341 & 05-03-2022 & 0.33 & 4.4$''\times$3.2$''$,75\degree & J0948+0022 & J1058+0133 \\
AS2COS0044.1 & 9:59:10.3 & +02:48:55.7 & 2.580 & 07-03-2022 & 0.20 & 4.3$''\times$3.1$''$,75\degree & J1016+0513 & J1058+0133\\
AS2COS0065.1 & 09:58:40.3 & +02:05:14.7 & 2.414 & 07-03-2022 & 0.20 & 2.8$''\times$2.0$''$,68\degree & J0948+0022 & J0854+2006\\
AS2COS0066.1 & 09:59:46.6 & +01:57:15.1 & 3.247 & 05-03-2022 & 0.46 & 3.1$''\times$2.6$''$,80\degree & J0948+0022 & J1058+0133\\
AS2UDS012.0 & 02:18:03.6 & $-$04:55:27.2 & 2.520 & 02-03-2022 & 0.26 & 2.6$''\times$2.0$''$,87\degree & J0228--0337 & J0238+1636\\
AS2UDS026.0 & 02:19:02.1 & $-$05:28:56.9 & 3.296 & 01-03-2022 & 0.40 & 3.2$''\times$2.6$''$,$-$78\degree & J0217--0820 & J0238+1636\\
AS2UDS126.0 & 02:15:46.7 & $-$05:18:49.2 & 2.436 & 02-03-2022 & 0.23 & 3.0$''\times$1.9$''$,72\degree & J0217--0820 & J0238+1636\\
& & & & 13--03--2022 & & & & J0423--0120\\
& & & & 22--03--2022 & & & & J2258--2758\\
\hline
\hline
\multicolumn{9}{l}{$^a$ For channels of 60 km s$^{-1}$.}
 \end{tabular}
\end{table*}

In this paper, we present ALMA Band 3 and 4 observations of the [\ion{C}{1}](1--0) emission line in a sample of 12 unlensed SMGs which are part of a new VLA survey of molecular gas in massive star-forming galaxies at high redshift \citep{friascastillo2023}. 
SMGs are dusty, high-infrared-luminosity ($L_\mathrm{IR} >$ 10$^{12}$ L$_\odot$; \citealt{magnelli2012lir_smg}) galaxies harboring some of the most intense starbursts that have ever occurred (SFR $\sim$100 -- 1000 M$_\odot$ yr$^{-1}$; \citealt{magnelli2012lir_smg,swinbank2014,dacunha2015,Dudzeviciute2020,AS2COSPEC_II2024}), fueled by large molecular gas reservoirs of 10$^{10}$ -- 10$^{11}$ M$_\odot$ \citep{greve2005,tacconi2008,bothwell2013,aravena2016,birkin2020,friascastillo2023}. Such luminous, massive galaxies are therefore the ideal targets to detect the faint CO(1--0) line with current facilities without the need for gravitational lensing. Combined with eight previously published [\ion{C}{1}](1--0) detections from galaxies in the parent sample \citep[][Huber A. in prep]{birkin2020,AS2COSPEC_II2024}, this makes a total of 20 high--redshift, unlensed SMGs with observation of the ground transition of both CO and [\ion{C}{1}].

This paper is structured as follows: we present the sample and ALMA observations, line emission and continuum detections in Section \ref{sec:Sample}. In Section \ref{sec:Analysis} we probe the relation between the [\ion{C}{1}] and infrared luminosities, as well as the relation between the CO(1--0) and [\ion{C}{1}](1--0) luminosities. We use line and line--to--continuum ratios to constrain the physical conditions in the ISM of our galaxies, and present a cross--calibration of [\ion{C}{1}] as a gas mass tracer through a comparison with CO(1--0) and dust continuum emission. We then model the effect of the CMB on line emission at high redshift. Finally, we present a discussion of our results in Section \ref{sec:Discussion} and our conclusions in Section \ref{sec:Conclusions}. Throughout this paper we assume a standard $\Lambda$CDM cosmology with $H_0$ = 67.8 km s$^{-1}$ Mpc$^{-1}$, $\Omega_\mathrm{M}$ = 0.310 and $\Omega_\Lambda$ = 0.690 \citep[][]{planck2016}.

\section{Data Reduction and Results} \label{sec:Sample}

\subsection{Sample}
We analyse [\ion{C}{1}](1--0) observations of 20 galaxies selected from an on--going \textit{Karl G. Jansky} Very Large Array (JVLA) survey, in which a total of 30 sources have been observed in CO(1--0) line emission in the K- and Ka-bands \citep{friascastillo2023}. The targets were selected from a sample of sources detected within 4 deg$^2$ of SCUBA-2 850$\mu$m imaging in the UKIDSS Ultra Deep Survey (UDS), Cosmological Evolution Survey (COSMOS), Chandra Deep Field North (CDFN), and Extended Groth Strip (EGS) fields from the S2CLS \citep{geach2017} and S2COSMOS \citep{simpson2019} surveys. The brightest submillimeter sources in these fields were subsequently followed up with ALMA (AS2UDS, \citealt{stach2018}; AS2COSMOS, \citealt{simpson2020}) and SMA \citep[EGS, CDFN,][]{hill2018} continuum imaging and further targeted with blind line scans using ALMA or NOEMA to obtain precise redshifts \citep[][Huber A. in prep]{birkin2020,AS2COSPEC2022,AS2COSPEC_II2024}. We refer the reader to \cite{friascastillo2023} for further details of the sample selection. Of the original parent sample, eight sources have published [\ion{C}{1}](1--0) emission line detections \citep{birkin2020,AS2COSPEC_II2024}, and we present new ALMA Band 3 and 4 observations for 12 more sources. Our sample thus comprises a total of 20 sources with [\ion{C}{1}](1--0) and CO(1--0) line observations.  

\subsection{JVLA CO(1--0) Data}

Observations of the CO(1--0) emission (rest-frame frequency: $\nu_\mathrm{rest}$ = 115.2712 GHz) in our sample were carried out with the JVLA between 2021 March and 2024 January (project 21A-254, P.I.: Hodge). The K- or Ka-band receivers were used in combination with the WIDAR correlator configured to 8-bit sampling mode to observe a contiguous bandwidth of 2 GHz (dual polarization) at 2MHz spectral resolution. Eight of the SMGs in this sample have CO(1--0) fluxes or upper limits reported in \cite{friascastillo2023}, so we take the line luminosities and recalculate the upper limits for consistency with our analysis (see Sec. \ref{sec:detections} for details on the flux extraction). For the rest of the sample, we calibrated the new data and obtained fluxes following \cite{friascastillo2023}. The final line luminosities are reported in Table \ref{tab:results}. 

\begin{sidewaystable*}
\small
\vspace{7cm}
\caption{Total [\ion{C}{1}](1--0) flux density, line width and line luminosity, gas masses inferred from [\ion{C}{1}](1--0) using $X_\mathrm{[CI]}=5.1\times10^{-5}$, CO(1--0) line luminosity, available mid--$J_\mathrm{up}$ CO transition and line luminosity, FIR luminosity and continuum flux density for the SMGs studied in this work. Sources below the solid line are published [\ion{C}{1}](1--0) observations from \cite[][and Huber A. in prep]{birkin2020,AS2COSPEC2022} that we include in this work with their corresponding new CO(1--0) observations.} \label{tab:results} 
\begin{adjustwidth}{-1.9cm}{}
\begin{tabular}{@{}ccccccccccc @{} }
 \hline 
 \hline
Target & $z$ & I$_\mathrm{[CI]}$ & FWHM$_\mathrm{[CI]}$ & L$'_\mathrm{[CI]}$ & $M_\mathrm{gas,[CI]}$ & $L'_\mathrm{CO(1-0)}$ & $J_\mathrm{up}$ & $L'_\mathrm{CO(J_{up}-(J_{up}-1))}$ & $L_\mathrm{IR}$ & $S_\mathrm{3mm}$  \\ 
& & [Jy km s$^{-1}$] & [km s$^{-1}$] & [10$^{10}$ K km s$^{-1}$ pc$^2$] & [10$^{10}$ $M_\odot$]& [10$^{10}$ K km s$^{-1}$ pc$^2$] & & [10$^{10}$ K km s$^{-1}$ pc$^2$] & [10$^{12}$ $L_\odot$]& [mJy] \\
\hline
AS2COS0001.1 & 4.625 & 0.54$\pm$0.20 & 680$\pm$170 & 2.44$\pm$0.91 & 9.9$\pm$3.7 &$<$19.8$^f$ & 5 & 2.7$\pm$0.6$^c$ & 36.3$^{+5.0}_{-6.7}$ & 0.38$\pm$0.04  \\
AS2COS0002.1 & 4.595 & 0.39$\pm$0.15 & -- & 1.75$\pm$0.67 & 7.1$\pm$2.7 &$<$20.8$^f$ & 5 & 3.8$\pm$0.8$^c$ & 9.5$^{+3.1}_{-2.9}$ & 0.22$\pm$0.03 \\
AS2COS0008.1 & 3.581 & 1.03$\pm$0.14 & 700$\pm$90 & 3.12$\pm$0.42 & 12.7$\pm$1.7 & 11.6$\pm$3.9$^b$ & 4 &5.2$\pm$0.6$^c$ & 15.9$^{+0.2}_{-1.1}$ & 0.40$\pm$0.03 \\
AS2COS0009.1 & 2.260 & $<0.42$ & 840$\pm$350 & $<$0.6 & $<$2.3 &$<$6.4$^b$ & 3 & 8.4$\pm$1.2$^c$  & 4.4$^{+0.8}_{-0.1}$ & 0.58$\pm$0.04 \\
AS2COS0014.1 & 2.921 & 0.76$\pm$0.35 & 750$\pm$160 & 1.65$\pm$0.76 &  6.7$\pm$3.1 & 9.1$\pm$3.4$^f$ & 3 & 9.1$\pm$1.5$^c$  & 10.7$^{+0.7}_{-1.0}$& 0.62$\pm$0.03 \\
AS2COS0023.1 & 4.341 & 0.70$\pm$0.14 &  480$\pm$80 & 2.87$\pm$0.57 & 11.6$\pm$2.3 & 7.0$\pm$3.1$^b$  & 4 & 9.1$\pm$0.7$^c$  & 20.4$^{+0.7}_{-0.5}$ & 0.21$\pm$0.03 \\
AS2COS0044.1 & 2.580 & $<$0.27 & -- & $<$0.47 & $<$2.0 & $<$13.3$^f$ & 3 & 5.0$\pm$0.6$^c$  & 3.1$^{+0.6}_{-0.7}$ &  1.34$\pm$0.02 \\
AS2COS0065.1 & 2.414 & 1.44$\pm$0.16 & 580$\pm$50 & 2.27$\pm$0.25& 9.2$\pm$1.0 & $<$13.2$^f$ & 3 & 8.1$\pm$0.8$^c$ & 9.3$^{+0.9}_{-0.7}$ & 0.46$\pm$0.02 \\
AS2COS0066.1$^*$ & 3.247 & 0.61$\pm$0.29 & 160$\pm$50 & 1.45$\pm$0.36 & 5.6$\pm$1.5 & 4.9$\pm$2.5$^f$  &  4 &3.7$\pm$0.4$^c$  & 6.8$^{+0.9}_{-0.9}$& 0.23$\pm$0.03 \\
AS2UDS012.0 & 2.520 & 0.72$\pm$0.26 &740$\pm$140 & 1.22$\pm$0.44 & 5.0$\pm$1.8 & 11.1$\pm$2.3$^b$ & 3 & 4.0$\pm$0.8$^d$ & 4.3$^{+0.1}_{-0.1}$ & 0.32$\pm$0.03 \\
AS2UDS026.0 & 3.296 & 0.99$\pm$0.31 & 700$\pm$150 & 2.63$\pm$0.82 & 10.6$\pm$3.3 & 6.3$\pm$2.9$^b$  & 4 &4.3$\pm$0.8$^d$  & 4.5$^{+0.8}_{-0.9}$ & 0.19$\pm$0.02 \\
AS2UDS126.0 & 2.436 & 1.17$\pm$0.19 & 680$\pm$70 & 1.87$\pm$0.30 & 7.6$\pm$1.2 &4.9$\pm$2.9$^b$ & 3 & 4.3$\pm$0.8$^d$ & 9.3$^{+2.6}_{-1.9}$ & 0.43$\pm$0.02 \\
\hline
AS2COS0011.1 & 4.78 & 1.1$\pm$0.3$^c$ & 640$\pm$60 & 5.1$\pm$1.7 & 20.7$\pm$6.9 & $<$35$^a$ & 5 & 6.7$\pm$0.8$^c$ & 12.6$^{+2.6}_{-2.3}$ & 0.47$\pm$0.02\\
AS2COS0031.1 & 3.64 & 1.0$\pm$0.2$^c$ & 400$\pm$60 & 3.1$\pm$0.7 & 12.6$\pm$2.8 & 12.7$\pm$3.5$^b$ &  4 & 8.9$\pm$0.7$^c$  & 14.8$^{+0.3}_{-0.3}$& 0.28$\pm$0.02\\
AS2UDS011.0 & 4.07 & 0.73$\pm$0.24$^d$ & 610$\pm$220 & 2.7$\pm$0.9& 10.9$\pm$3.6 &4.7$\pm$2.7$^b$ &  4 & 3.9$\pm$1.0$^d$ & 11.8$^{+1.6}_{-2.0}$ & 0.12$\pm$0.02 \\
AS2UDS014.0$^{**}$ & 3.8 & 0.23$\pm$0.36$^d$ & 1560$\pm$5 & $<3.6$& $<$14.6& $<$15.2$^b$  &  4 & 9.9$\pm$1.2$^d$  & 8.7$^{+1.9}_{-0.2}$ &0.19$\pm$0.03 \\
CDFN2 & 4.42 &0.90$\pm$0.10$^e$& 500$\pm$50&3.79$\pm$0.4& 15.4$\pm$1.6 &12.9$\pm$3.8$^b$ &4& 8.2$\pm$0.3$^e$ & 17.0$^{+0.8}_{-0.5}$ & 0.35$\pm$0.03 \\
CDFN8 & 4.14 &0.61$\pm$0.12$^e$& 650$\pm$70 &2.32$\pm$0.44& 9.4$\pm$1.8 &7.0$\pm$4.3$^b$&4& 4.6$\pm$0.3$^e$ & 8.0$^{+1.8}_{-1.2}$&0.20$\pm$0.04\\
AEG2 & 3.67 &0.70$\pm$0.13$^e$& 830$\pm$80 &2.21$\pm$0.39& 9.0$\pm$1.6 &11.4$\pm$3.4$^b$ &4& 4.4$\pm$0.2$^e$& 5.8$^{+0.1}_{-0.1}$&0.23$\pm$0.04\\
AEG3 & 4.05 &0.35$\pm$0.07$^e$& 600$\pm$70&1.29$\pm$0.22& 5.2$\pm$0.9 &$<$6.0$^b$&4& 2.7$\pm$0.2$^e$& 6.9$^{+2.7}_{-1.2}$&0.11$\pm$0.03\\
\hline
\hline
\multicolumn{11}{l}{$a$ We downloaded and calibrated the data from the NRAO archive, project 21A-254 (P.I.: Hodge). Fluxes were calculated following \cite{friascastillo2023}.} \\
\multicolumn{11}{l}{Line luminosities taken from: $b$) \cite{friascastillo2023}; $c$) \cite{AS2COSPEC_II2024}; $d$) \cite{birkin2020}, $e$) Huber A. in prep., $f$) Hodge, J. priv.comm.}\\
\multicolumn{11}{l}{$^*$ AS2COS0066.1 falls on the edge of the ALMA Band, so we only cover the velocity range spanned by the main component of the emission seen in }\\
\multicolumn{11}{l}{CO(4--3) \citep{AS2COSPEC2022}. This component represents $\sim$75\% of the total flux, so we scale the inferred CO(1--0) luminosity down by 25\%, and note that this}\\
\multicolumn{11}{l}{measurement is a lower limit of the total [\ion{C}{1}] line flux.} \\ 
\multicolumn{11}{l}{$^{**}$ Although AS2UDS014.0 is reported as a detection in \cite{birkin2020}, it does not pass our 2$\sigma$ criterion for a detection. Therefore, we take the three times the reported error on the} \\
\multicolumn{11}{l}{line luminosity as an upper limit for this source.} \\
\end{tabular}
\end{adjustwidth}{}
\end{sidewaystable*}

\begin{figure*}[!ht]
    \vspace{-0.7cm}
    \hspace{-1.25cm}
    \includegraphics[scale=0.6]{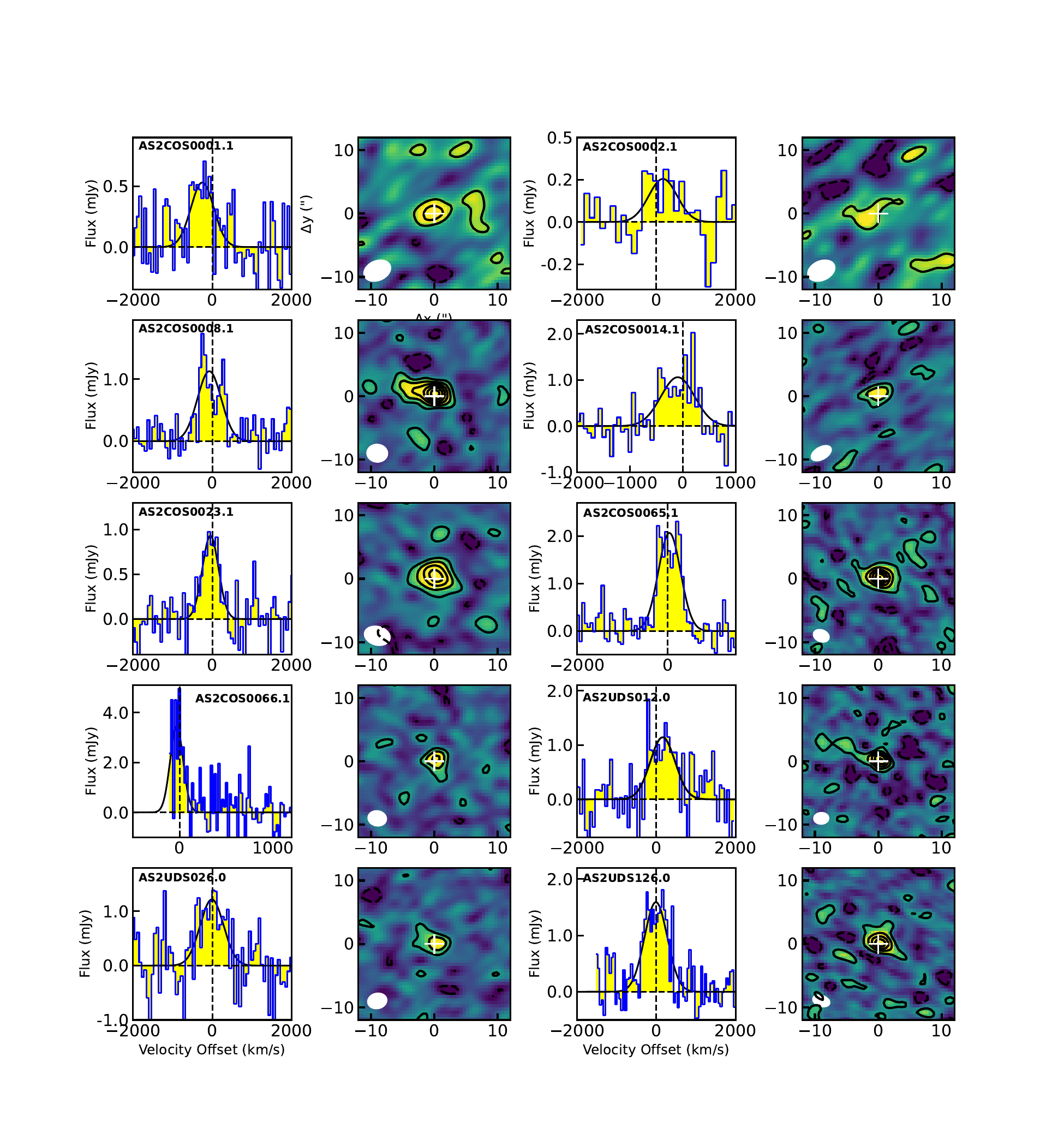}
    \caption{[\ion{C}{1}](1--0) line emission for the detections in our sample of SMGs. The spectra (blue line and yellow fill, left panels) are extracted within a 2.5$''$ radius aperture to maximise the SNR. The spectra were fit with a single Gaussian model allowing for a varying line width, shown by the black curve. The 0th-moment maps (right panels) were collapsed over a velocity range equal to the FWHM of the respective mid--$J_\mathrm{up}$ CO emission line and show a 20$''\times$20$''$ field of view. The systemic velocity is based on the redshift derived from the mid--$J_\mathrm{up}$ CO lines. The white cross indicates the peak of the mid--$J_\mathrm{up}$ CO line emission. Contours start at 2$\sigma$ and increase in steps of 2$\sigma$, except for AS2COS0065.1 and AS2UDS126.0, where they do so in steps of 4$\sigma$. The white ellipse shows the FWHM of the beam for each source.}
    \label{detections}
\end{figure*}

\subsection{ALMA Data and Reduction}

We obtained ALMA Band 3 and 4 observations during Cycle 8 (Project ID: 2021.1.01342.S). Data were collected in configurations C43--1/C43--2 between March 1 and 22 2022.

We adopt calibrations performed in the second level of quality assurance (QA2). Flagging and calibrations were done using \textsc{CASA} version 6.2.1.7 \citep[pipeline version: 2021.2.0.128,][]{casa}, which was also used for imaging. We first subtracted the continuum using the task \textsc{uvcontsub} in \textsc{CASA} after excluding the channels containing line emission. A priori, we used the line widths of the mid--$J_\mathrm{up}$ CO line emission reported in \cite{birkin2020} and \cite{AS2COSPEC_II2024} and masked channels within $\pm$2$\times$ the Full Width Half Maximum (FWHM) of the expected line, centered on the expected frequency of the [\ion{C}{1}](1--0) line. We then imaged the line--free channels in the continuum--subtracted image cubes to verify that all emission had been fully removed. No further continuum subtraction was necessary.

The continuum--subtracted visibilities were then imaged  and cleaned using the \textsc{tclean} algorithm in \textsc{CASA}, with a cleaning threshold of 2$\sigma$. We adopted natural weighting to maximise the signal-to-noise ratio (SNR) of the detections, which resulted in final beam sizes ranging from 2.6$''$ to 4.7$''$ (Table \ref{tab:sample}) at FWHM. At this resolution, the targeted galaxies are not (or only marginally) spatially resolved. We binned the data cubes spectrally to channels of 60 km s$^{-1}$ to increase the SNR. The resulting data cubes reach a noise level of 0.2--0.5 mJy beam$^{-1}$ per channel (see Table \ref{tab:sample}), estimated over the emission--free region of the channels. 

\subsection{[\ion{C}{1}] Line and Continuum Detections} \label{sec:detections}

We began by collapsing the cleaned line cubes over the velocity width of the corresponding  mid--$J_\mathrm{up}$ CO line from \cite{birkin2020} or \cite{AS2COSPEC_II2024}. This allows us to initially assess whether there is a spatial offset between the [\ion{C}{1}] emission and the coordinates given in Table \ref{tab:sample}, derived from the ALMA and SMA continuum imaging \citep[][Huber A. in prep.]{stach2018,simpson2020}. Only AS2COS0001.1 and AS2COS0002.1 show an offset of $\sim$0.5$''$. This offset might be caused by the low SNR of the emission in AS2COS0002.1. In the case of AS2COS0001.1, the source has a known companion only a few arcseconds away \citep{AS2COSPEC2022}, which we do not resolve due to the resolution of our observations and might skew the position of the peak emission. We extract the line spectra from the position of the peak emission using an aperture of 5$''$ in diameter, to maximise the SNR. We detect [\ion{C}{1}](1--0) emission in 10 out of the 12 sources in the new [\ion{C}{1}](1--0) sample. The undetected sources, AS2COS0009.1 and AS2COS0044.1, are amongst the faintest far--infrared sources in the sample, which may explain their non--detections (Table~\ref{tab:results}). For the detected sources, the FWHM was derived from a single--Gaussian fitting to the spectra shown in Fig.~\ref{detections}.

To avoid any bias due to line structure and to be consistent with the analysis performed for CO(1--0) in \cite{friascastillo2023}, we derive line fluxes using the intensity-weighted (0$^\mathrm{th}$--moment) maps collapsed over a velocity range twice the corresponding mid--$J_\mathrm{up}$ CO line width for each source \citep{birkin2020,AS2COSPEC_II2024}:

\begin{equation}
    M_0 = I_\mathrm{CO} = \int I_\nu d\nu.
\end{equation}

The median line-width ratio of CO(4--3) or CO(3--2) to [\ion{C}{1}](1--0) is 1.0 $\pm$ 0.1, and therefore the choice of FWHM (CO or [\ion{C}{1}]) does not systematically impact the final recovered flux. Additionally, for the sources for which the SNR of the CO(1--0) data allows the fitting of a Gaussian profile to derive the FWHM, both the [\ion{C}{1}](1--0) and CO(1--0) emission lines show consistent FWHM (FWHM$_\mathrm{[CI](1-0)}$/FWHM$_\mathrm{CO(1-0)}=$1.03$\pm$0.3). 

In order to determine the optimal aperture to obtain the line fluxes, we perform a curve-of-growth analysis on these 0$^\mathrm{th}$--moment maps by extracting flux densities from a set of circular apertures of increasing diameter, from 1.5$''$ to 40$''$. We find that an aperture of diameter 2$\times$ the synthesised beam FWHM recovers most of the line flux while maximising the SNR, so we adopt that aperture size to extract final fluxes. This is consistent with the fluxes obtained using the \textsc{CASA} task \textsc{imfit}, which conducts 2D Gaussian fitting on the 0$^\mathrm{th}$--moment maps. For the two sources without detections, AS2COS0009.1 and AS2COS0044.1, we place 3$\sigma$ upper limits on the emission by calculating the RMS over an aperture covering a region of the 0$^\mathrm{th}$--moment map away from the position of the target, and then correcting for an aperture of 2$\times$ the synthesised beam size. We extract continuum fluxes in a similar manner by using a circular aperture of size 2$\times$ the synthesised beam size on the continuum images. The final fluxes and FWHM are compiled in Table \ref{tab:results}.\footnote{AS2COS0001.1 and AS2COS0008.1 have companion galaxies nearby \citep{AS2COSPEC2022} which we cannot resolve at the resolution of our observations. Although the bulk of the emission is most likely dominated by the primary source, the [\ion{C}{1}] line flux densities are likely overestimated. Since the CO(1--0) measured fluxes also suffer from the same problem, we do not correct the fluxes presented and treat both sources as a single target. }

We convert the [\ion{C}{1}] line intensities into line luminosities following \cite{solomon2005}:
\begin{equation}
    L'_\mathrm{[CI](1-0)} = 3.25 \times 10^7 \ I_\mathrm{[CI]} \ \mathrm{\nu{^{-2}_{obs}}} \ D\mathrm{{^2_{L}}} \ (1+\mathrm{z})^{-3} \ \mathrm{K \ km \  s^{-1} \ pc^2} ,
\end{equation}
\noindent where $I_\mathrm{[CI]}$ is the integrated line flux from the 0$^\mathrm{th}$--moment map in Jy km s$^{-1}$, $\mathrm{\nu_{\text{obs}}}$ is the observed frequency in GHz and $D\mathrm{_{\text{L}}}$ is the luminosity distance in Mpc. We derive line luminosities in the range $<$0.47--3.1$\times$10$^{10}$ K km s$^{-1}$ \ pc$^2$. Combined with the eight sources with published [CI](1--0) luminosities, our total sample spans the luminosity range $<$0.47--5.1$\times$10$^{10}$ K km s$^{-1}$ \ pc$^2$ (see Table \ref{tab:results}).

\section{Analysis} \label{sec:Analysis}

\subsection{[\ion{C}{1}] Line Luminosities}

\begin{figure*}[!ht]
    \centering
    \includegraphics[scale=0.4]{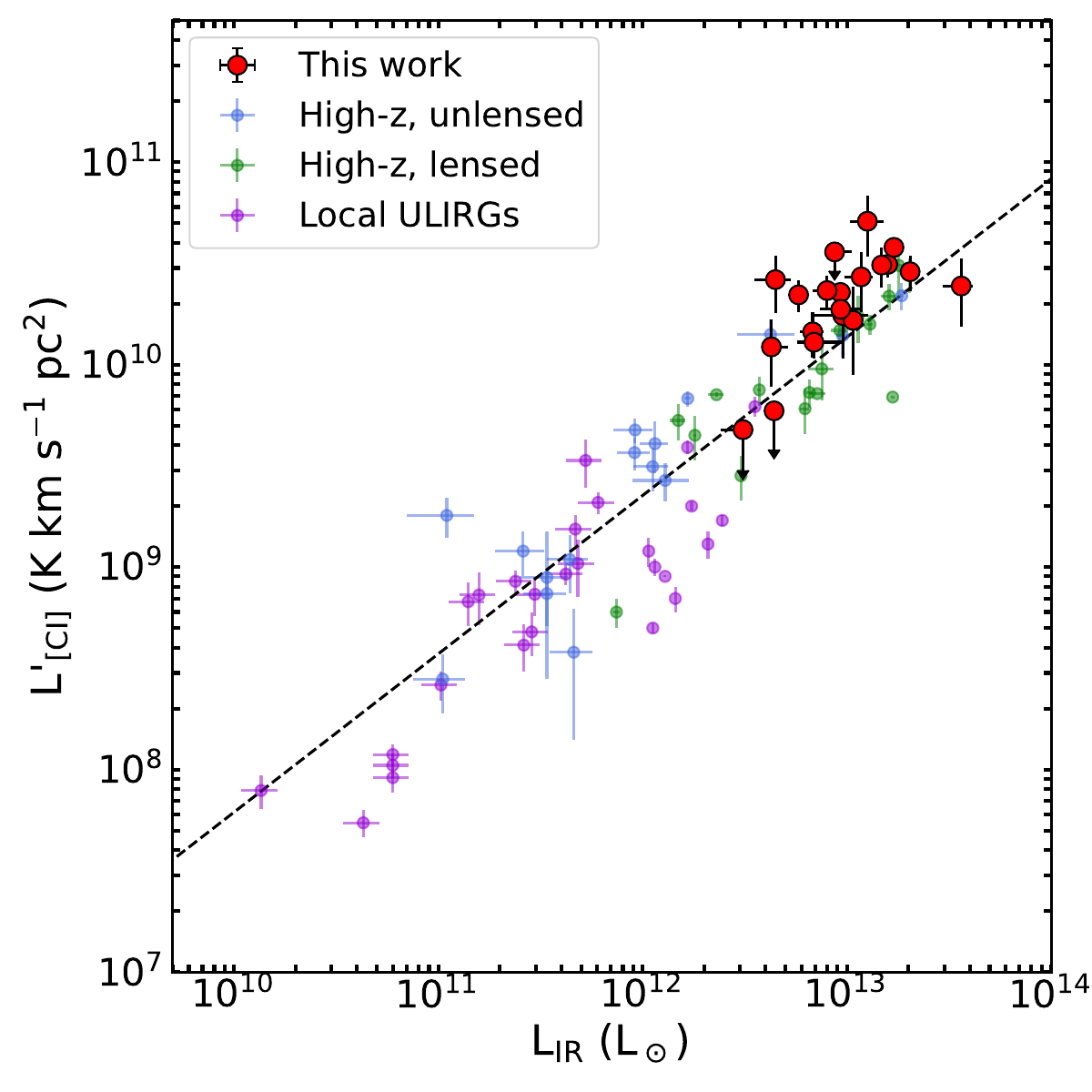}
    \includegraphics[scale=0.4]{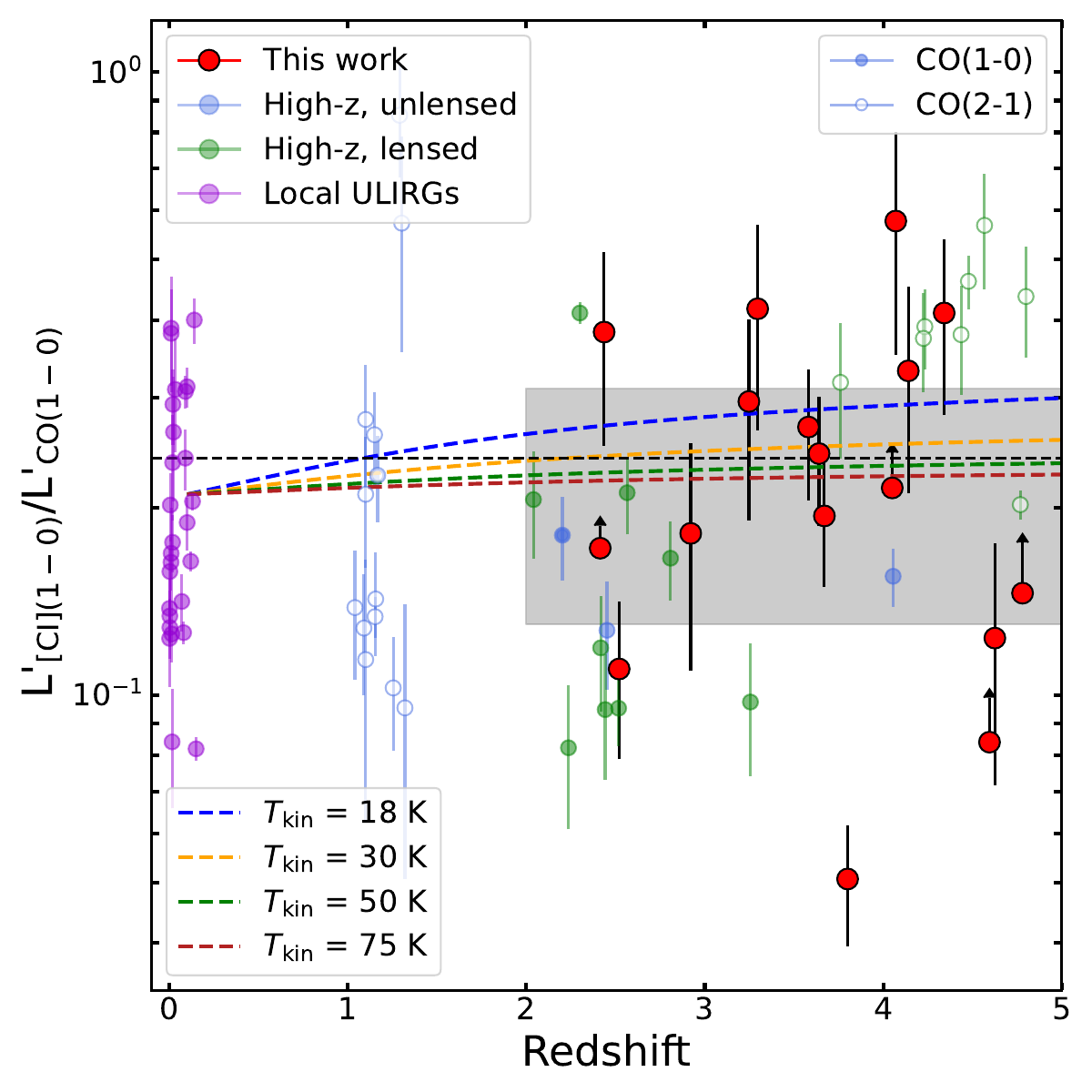}
    \caption{\textit{Left}:[\ion{C}{1}](1--0) line luminosities compared to $L_\mathrm{IR}$ for our sample. As comparison, we have added the local ULIRGs from \cite{liu2019}, \cite{kamenetzky2016} and \cite{montoya2023}; the $z\sim$1 main sequence galaxies from \cite{valentino2018}, \cite{popping2017}, \cite{jin2019} and \cite{boogaard2020}; and the SMGs from \cite{walter2011,alaghband2013,bothwell2017,Canameras2018,nesvadba2019,harrington2018,dannerbauer2019}. We exclude galaxies classified as AGN according to \cite{valentino2020}. Literature luminosities are corrected for lensing magnification where necessary. The dashed line shows the best--fit relation from \citet{valentino2018}. \textit{Right}: $L'_\mathrm{[CI](1-0)}$/$L'_\mathrm{\text{CO(1-0)}}$ ratio as a function of redshift. The dashed black line shows the median value for our sample, log($L'_\mathrm{[CI](1-0)}$/$L'_\mathrm{\text{CO(1-0)}}$) = $-$0.72, and the shaded region marks the absolute deviation around the median (0.14 dex). The dashed colored lines show the expected change in the ratio with redshift due to the warmer CMB for $n_\mathrm{H_2}=10^2$ cm$^{-3}$ and a range of gas kinetic temperatures (see Section~\ref{sec:CMB}).}
    \label{lci_lum}
\end{figure*}

First, we put the [\ion{C}{1}] luminosities of the galaxies in our sample in the context of other low--and high--redshift observations. Fig.~\ref{lci_lum} (left) shows the $L'_\mathrm{[CI](1-0)}$ line luminosities as a function of infrared luminosity, $L_\mathrm{IR}$. $L_\mathrm{IR}$ was estimated through SED fitting with \textsc{MAGPHYS} \citep{highz_magphys}, integrating over 8--1000$\mu$m. For details of the photometry used, we refer the reader to \cite{simpson2020} for sources in AS2COSMOS, \cite{Dudzeviciute2020} for AS2UDS and \cite{friascastillo2023} for sources in the CDFN and EGS fields. If we take $L'_\mathrm{[CI](1-0)}$ and $L_\mathrm{IR}$ as proxies for gas mass and SFR, respectively, their relation is equivalent to the spatially--integrated Schmidt--Kennicutt relation \citep{schmidt1959,kennicutt1998,kennicutt+evans2012}. We compare them with the local ULIRGs from \cite{liu2019}, \cite{kamenetzky2016} and \cite{montoya2023}; the $z\sim$1 MS galaxies from \cite{valentino2018} and \cite{boogaard2020}; and the SMGs from \cite{walter2011,danielson2011,alaghband2013,bothwell2017,Canameras2018,harrington2018,nesvadba2019,dannerbauer2019}. We exclude galaxies classified as AGN according to \cite{valentino2020}, and we restrict our compilation to only sources with either CO(1--0) or CO(2--1) observations to minimise the uncertainty introduced by excitation corrections. The literature data with mid--$J_\mathrm{up}$ CO observations has been converted to CO(1--0) using $r_\mathrm{21}$ = $L'_\mathrm{\text{CO(2-1)}}$/$L'_\mathrm{\text{CO(1-0)}}$ = 0.9 \citep{birkin2020}. When necessary, fluxes and luminosities have been corrected for magnification. Although the SMGs in this work are marginally above the $L'_\mathrm{[CI](1-0)}$--$L_\mathrm{IR}$ relation found for other ULIRGs and main--sequence galaxies \citep{valentino2018}, they are mostly within the 0.26 dex scatter of the relation (we assess whether our sample could be drawn from the $L'_\mathrm{[CI](1-0)}$--$L_\mathrm{IR}$ relation from \cite{valentino2018} using a $\chi^2$ test and find a $p$--value = 0.7, consistent with being drawn from the same distribution).

As shown in Fig.~\ref{lci_lum} (left) and Table \ref{tab:results}, we do not detect [\ion{C}{1}] emission in two sources, AS2COS0009.1 and AS2COS00044.1.
Their 3$\sigma$ upper limits fall within the $L'_\mathrm{[C1]](1-0)}$--$L_\mathrm{IR}$ relation, so it is possible that they are too gas--poor to be detected given the sensitivity of our data.

The right panel of Fig. \ref{lci_lum} shows the $L'_\mathrm{[CI]](1-0)}$/$L'_\mathrm{\text{CO(1-0)}}$ ratio as a function of redshift. We calculate a median ratio of log($L'_\mathrm{[CI](1-0)}$/$L'_\mathrm{\text{CO(1-0)}}$) = $-$0.62$\pm$0.14, comparable with the value for the high--redshift SMG literature sample and the local ULIRGs ($-$0.74$\pm$0.12). This ratio is also consistent with the value of log($L'_\mathrm{[CI](1-0)}$/$L'_\mathrm{\text{CO(2--1)}}$) = ($-$0.69$\pm$0.04) found by \cite{valentino2018}. The $z\sim1$ main sequence galaxies have a lower median value ($-$0.85$\pm$0.16), although it is still consistent with our sample. We also find no difference in $L'_\mathrm{[CI]](1-0)}$/$L'_\mathrm{\text{CO(1-0)}}$ ratios when compared to total $L_\mathrm{IR}$. Therefore, [\ion{C}{1}](1--0) and CO(1--0) appear to be correlated on global scales regardless of galaxy type, $L_\mathrm{IR}$ and redshift.

\subsection{PDR Modelling} \label{sec:PDR}

\begin{figure}
    \centering
    \includegraphics[width=\linewidth]{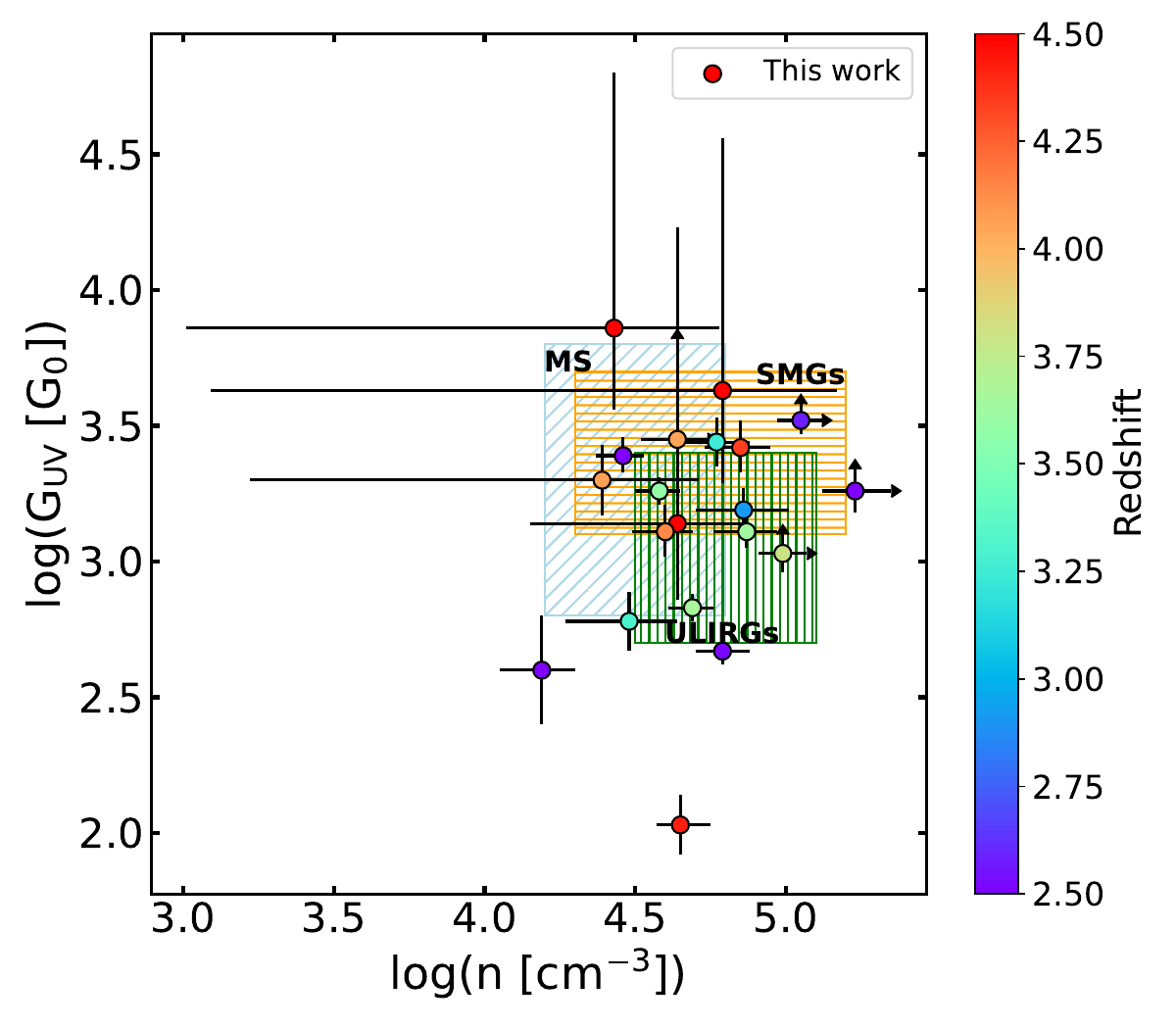}
    \caption{ISM density ($n$) and UV radiation field (G$_\mathrm{UV}$) derived from PDR modelling for the sample in this work. Our sources are denoted by the circles and color coded by redshift. We compare to the literature sample from Fig.~\ref{lci_lum} with available [\ion{C}{1}](1--0), mid$-J_\mathrm{up}$ CO and FIR data. Overall, our galaxies span the same parameter space as other sources at high redshift, with a median $n=$10$^{4.7\pm0.2}$ cm$^{-3}$ and a median UV radiation field of G$_\mathrm{UV}$ = 10$^{3.2\pm0.2}$ G$_0$. The large scatter in G$_\mathrm{UV}$ likely reflects the varied nature of the sources studied in this work. }
    \label{pdr}\label{fig:pdr}
\end{figure}

How do the physical conditions (density, UV irradiation) of the interstellar medium in our sample compare to other high- and low-redshift galaxies? To answer this question, we use photo-dissociation region (PDR) modelling of the CO, [\ion{C}{1}] and far-infrared continuum data available for our sources. For our analysis, we use the \textsc{PDRToolbox} \citep{kaufman2006,pound2008} suite of one-dimensional, semi--infinite slab models with constant density. These models solve simultaneously for the chemistry, thermal balance, and radiative transfer, assuming metal, dust, and polycyclic aromatic hydrocarbons abundances, and a gas microturbulent velocity dispersion. For each combination of properties, every model is described in terms of the (number) density of H nuclei ($n$ [cm$^{-3}$]) and the intensity of the incident far--ultraviolet radiation ($G_\mathrm{UV}$, in units of the local interstellar field $G_0=1.6\times10^{-3}$ erg cm$^{-2}$ s$^{-1}$, \citealt{hanbing1968}). 

For the PDR modelling, we consider the [\ion{C}{1}](1--0), CO(1--0), mid-$J$ CO and $L_\mathrm{FIR}$.
The CMB heating is not implemented in the \textsc{PDRToolbox}. We compare the data to default, solar-metallicity \textsc{PDRToolbox} models - we consider them to be a good ``default'' scenario for dusty SMGs (see Section~
\ref{sec:Discussion}). As the \textsc{PDRToolbox} models consider only a single illuminated cloud face, we need to implement corrections for the optically thin tracers (in our case, FIR continuum and [\ion{C}{1}] which will emit from both the near and far side of the clouds, whereas optically thick tracers (CO) will only be seen from the near side of the cloud. Namely, for the FIR continuum and [\ion{C}{1}], we multiply the predicted fluxes by a factor of 2.

The [\ion{C}{1}](1--0)/$L_\mathrm{FIR}$ and CO(1--0)/$L_\mathrm{FIR}$ ratio, with [\ion{C}{1}] and CO tracing the extended gas mass and $L_\mathrm{FIR}$ tracing the SFR, allows us to constrain G$_\mathrm{UV}$. On the other hand, the ratio of [\ion{C}{1}](1--0) or CO(1--0) and a mid--$J_\mathrm{up}$ CO line effectively traces the gas density, while being almost insensitive to $G_\mathrm{UV}$. We thus use a combination of these ratios to determine the PDR properties for this sample of SMGs. Where appropriate, we use the 3$\sigma$ upper/lower limits. We repeat the process excluding the CO(1--0) line: the results do not change appreciably except for AS2COS0009.1 ($G_\mathrm{UV}$ increases by 0.8~dex), AS2UDS12.0 ($n$ increases by 0.5~dex) and AS2UDS126.0 ($G_\mathrm{UV}$ increases by $\sim$2~dex). For a consistent comparison, we take the luminosities of the comparison sample and calculate the densities and radiation field strengths.

\begin{table}[!htb]
\centering
\caption{Results of PDR modelling for individual sources. We list inferred $G_\mathrm{UV}$ and $n_\mathrm{H_2}$ values
}\label{tab:pdr}
\begin{tabular}{@{}l|cc @{}}
 \hline \hline
Target & $G_\mathrm{UV}$ & $n_\mathrm{H_2}$\\
& [log $G_0$] & [log cm$^{-3}$] \\
\hline
AS2COS0001.1	&	$3.86^{+0.94}_{-0.30}$	&	$4.43^{+0.35}_{-1.42}$ \\
AS2COS0002.1    &	$3.63^{+0.93}_{-0.34}$	&	$4.79^{+0.38}_{-1.70}$ \\ 
AS2COS0008.1    &	$3.26^{+0.05}_{-0.05}$	&	$4.58^{+0.07}_{-0.08}$ \\ 
AS2COS0009.1	&	$3.26^{+0.04}_{-0.05}$	&	$5.23^{+0.12}_{-0.11}$ \\ 
AS2COS0014.1	&	$3.19^{+0.08}_{-0.09}$	&	$4.86^{+0.15}_{-0.16}$ \\
AS2COS0023.1	&	$3.42^{+0.10}_{-0.09}$	&	$4.85^{+0.10}_{-0.12}$ \\ 
AS2COS0044.1	&	$3.52^{+0.06}_{-0.05}$	&	$5.05^{+0.07}_{-0.08}$ \\
AS2COS0065.1	&	$3.39^{+0.07}_{-0.06}$	&	$4.46^{+0.07}_{-0.09}$ \\ 
AS2COS0066.1	&	$3.44^{+0.09}_{-0.09}$	&	$4.77^{+0.11}_{-0.13}$ \\ 
AS2UDS012.0	    &	$2.67^{+0.05}_{-0.05}$	&	$4.79^{+0.09}_{-0.09}$ \\ 
AS2UDS026.0	    &	$2.78^{+0.11}_{-0.11}$	&	$4.48^{+0.16}_{-0.21}$ \\ 
AS2UDS126.0		&	$2.60^{+0.20}_{-0.20}$	&	$4.19^{+0.11}_{-0.14}$ \\ 
\hline
AS2COS0011.1	&	$3.14^{+1.09}_{-0.28}$	&	$4.64^{+0.23}_{-0.49}$ \\
AS2COS0031.1	&	$3.11^{+0.06}_{-0.06}$	&	$4.87^{+0.10}_{-0.11}$ \\ 
AS2UDS011.0	    &	$3.30^{+0.13}_{-0.13}$	&	$4.39^{+0.19}_{-0.29}$ \\ 
AS2UDS014.0			&	$3.03^{+0.07}_{-0.07}$	&	$4.99^{+0.08}_{-0.08}$ \\ 
CDFN2	&	$2.03^{+0.11}_{-0.11}$	&	$4.65^{+0.07}_{-0.08}$ \\ 
CDFN8	&	$3.11^{+0.10}_{-0.09}$	&	$4.60^{+0.09}_{-0.11}$ \\ 
AEG2	&	$2.83^{+0.05}_{-0.05}$	&	$4.69^{+0.07}_{-0.08}$ \\ 
AEG3	&	$3.45^{+0.37}_{-0.18}$	&	$4.64^{+0.10}_{-0.12}$ \\ 

\hline
\hline
 \end{tabular}
\end{table}

We show the results of the PDR modelling in Fig.~\ref{pdr} and Table~\ref{tab:pdr}. The SMGs show fairly high densities, with a median $n=$10$^{4.7\pm0.2}$ cm$^{-3}$. We refit the literature sample of MS and local galaxies from \cite{valentino2020}, and high--redshift SMGs \citep{alaghband2013,bothwell2017} for consistency with our models. The UV radiation field has a median value of $G_\mathrm{UV}$ = 10$^{3.2\pm0.2}$ $G_0$. While $G_{UV}$ and $n$ likely vary significantly across individual galaxies (see, e.g., the resolved study of SDP.81 by \citealt{rybak2020}), overall, SMGs in our sample occupy a similar parameter space as the main--sequence galaxies, as well as other high--redshift SMGs. The large spread in $G_\mathrm{UV}$ reflects the varied nature of the SMGs in our sample, as indicated by the wide range of CO excitation (i.e., CO(3--2)/CO(1--0) ratios spanning more than 1~dex; \citealt{friascastillo2023}).


\subsection{Comparison Between Molecular Gas Tracers} \label{comparison_co_ci}

We now cross--calibrate CO(1--0), [\ion{C}{1}](1--0) and dust continuum as gas mass tracers and test their agreement. Fig. \ref{lci_lum} (right) shows that the SMGs in this work follow the correlation between CO(1--0) and [\ion{C}{1}](1--0) line luminosities seen at low redshift \citep[e.g.,][]{jiao2017,jiao2019,montoya2023}, at least when averaged over global scales. This relation, along with the similarity in FWHM of the line profiles, suggests that the line emission from both lines arises from similar volumes in the galaxy and has been one of the main arguments for the use of [\ion{C}{1}] as a molecular gas tracer. 

The absolute values of the light--to--mass conversion factors for [\ion{C}{1}] and CO(1--0) have been the subject of extensive study in the literature. For [\ion{C}{1}], the conversion is directly dependent on the assumed carbon abundance, $X_\mathrm{[CI]}$, which varies with metallicity \citep{heintz_watson2020}. $X_\mathrm{[CI]}$ estimates in the literature range from $\sim1-2\times10^{-5}$ for MS galaxies \citep{valentino2018,jiao2019,boogaard2020} to $\sim4-8\times10^{-5}$ for ULIRGs and SMGs \citep{walter2011,alaghband2013,bothwell2017,jiao2017,gururajan2023}. CO(1--0), on the other hand, relies on the choice of $\alpha_\mathrm{CO}$, which also depends on metallicity, turbulence and gas kinematics \citep{bolatto2013}. Empirical calibrations for $\alpha_\mathrm{CO}$ derive values from $\sim0.8-1$ M$_\odot$ (K km s$^{-1}$ pc$^2$)$^{-1}$ for local ULIRGs and some SMGs \citep[e.g.,][]{downes1998,danielson2011,papadopoulos2012,chen2017,calistro-rivera2018,birkin2020,amvrosiadis2023} to $\sim3-5$ M$_\odot$ (K km s$^{-1}$ pc$^2$)$^{-1}$ for local discs and main sequence galaxies at high redshift \citep{sandstrom2013,remy2017,cormier2018,dunne2022}.

\subsubsection{[\ion{C}{1}](1--0) Line Emission as a Gas Mass Tracer} \label{sec:CI}

\begin{figure*}[!hbt]
    \centering
    \includegraphics[scale=0.4]{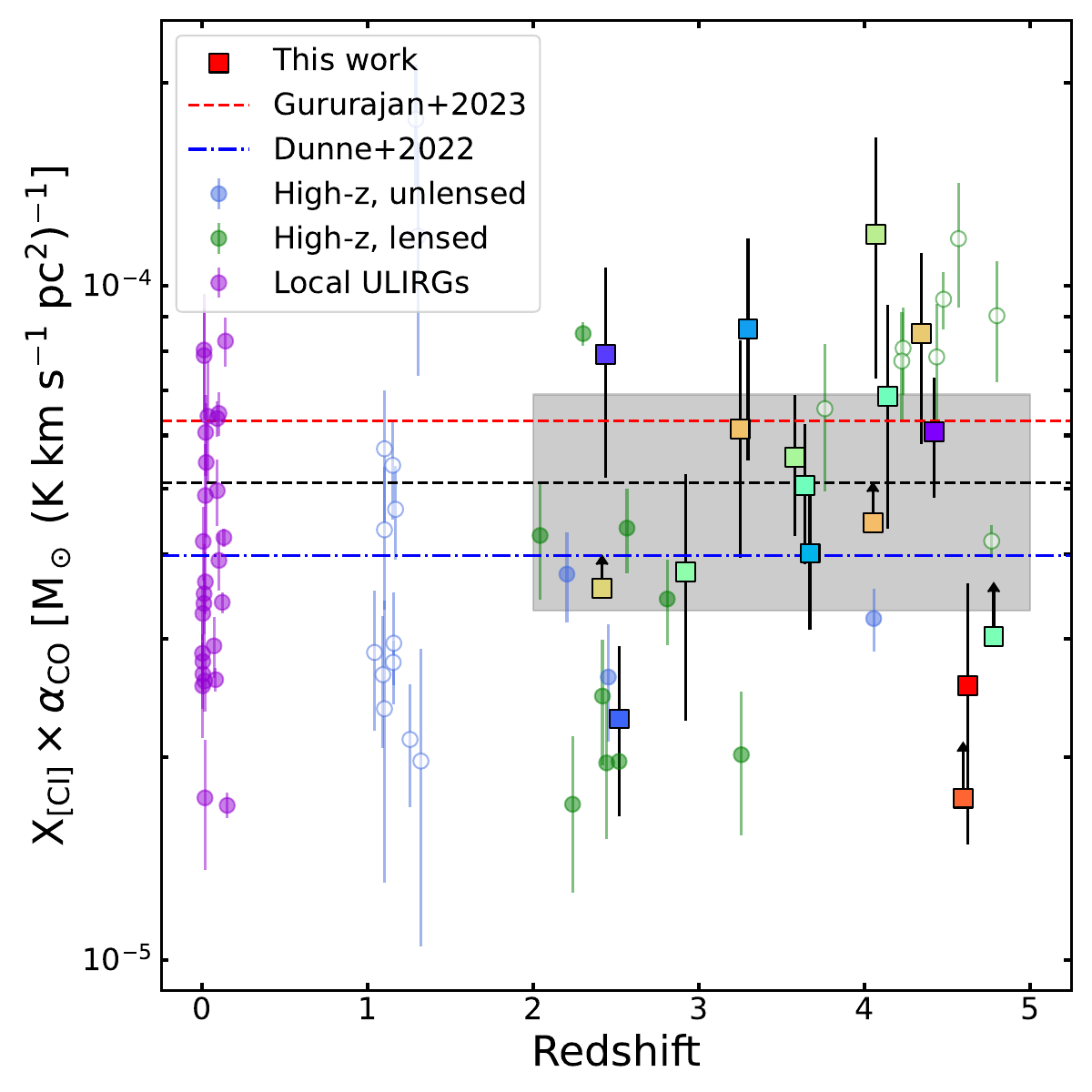}
    \includegraphics[scale=0.4]{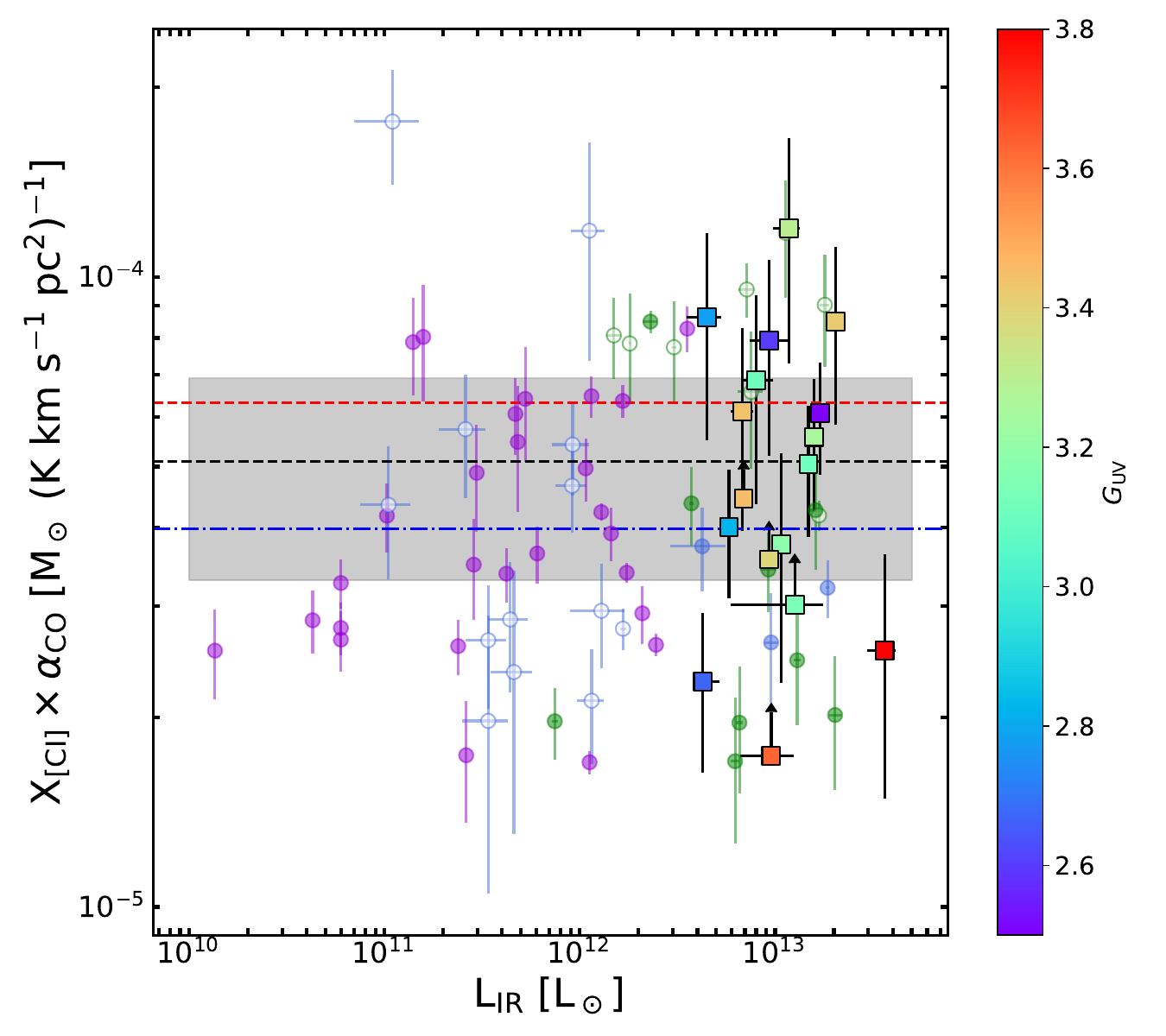}
    \caption{ $X_\mathrm{[CI]}\times\alpha_\mathrm{CO}$ factor as a function of redshift (left) and L$_\mathrm{IR}$ (right) for our sample. Our sample is shown in squares, color coded by their FUV radiation field. The literature compilation is as indicated in Fig.~\ref{lci_lum}. The mean value for the sample in this work is shown as a black dashed line, and the gray-shaded region represents the 1$\sigma$ standard deviation around the sample mean, (4.5$\pm$2.0)$\times$10$^{-5}$ (K km s$^{-1}$ pc$^2$)$^{-1}$. The dashed lines mark the mean values from \cite{gururajan2023} and \cite{dunne2022}. We fit the full sample to look for trends with either parameter, but do not find any dependence of $X_\mathrm{[CI]}\times\alpha_\mathrm{CO}$ on either redshift or L$_\mathrm{IR}$. The lack of correlation with $G_\mathrm{UV}$ suggests that $X_\mathrm{[CI]}\times\alpha_\mathrm{CO}$ is also insensitive to local conditions of the ISM in our sample.}
    \label{conversion_factor}
\end{figure*}

Following \cite{papadopoulos2004} and \cite{dunne2022}, we calculate the total molecular gas mass from the [\ion{C}{1}] line luminosity using the following expression:

\begin{equation} \label{ci_equation}
    \begin{aligned}
         M_\mathrm{H_2}^\mathrm{[CI]} = & \frac{0.0127}{X_{\mathrm{[CI]}} \ Q_{10}} \ \biggl(\frac{D_L^2}{1+z}\biggl) \ I_{\mathrm{[CI](1-0)}} \ \ M_\odot,
    \end{aligned}
\end{equation}

or in terms of line luminosity:

\begin{equation} \label{ci_gas}
M_\mathrm{H_2}^\mathrm{[CI]} = \frac{9.51 \times 10^{-5}}{X_{\mathrm{[CI]}} \ Q_{10}} \ L'_{\mathrm{[CI}](1-0)} \ M_\odot,
\end{equation}

where $X_\mathrm{[CI]}$ is the C/H$_2$ abundance ratio and $Q_\mathrm{10}$ is the [\ion{C}{1}] excitation factor of the [\ion{C}{1}] $J$ = 1 $\rightarrow$ 0 level. The value of $Q_\mathrm{10}$ is straightforward to calculate under the assumption of local thermodynamic equilibrium (LTE):

\begin{equation}
    Q(T_\mathrm{ex}) = 3e^{(-T_1/T_\mathrm{ex})} / (1 + 3e^{(-T_1/T_\mathrm{ex})} + 5e^{(-T_2/T_\mathrm{ex})}),
\end{equation}

\noindent where $T_1$ = 23.6 K and $T_2$ = 62.5 K are the excitation energy levels of atomic carbon. We adopt a typical value of $T_\mathrm{ex}$ = 30 K \citep{weiss2003,walter2011,harrington2021}, which results in $Q_\mathrm{10}$ = 0.46. We note that, above $T_\mathrm{ex}\sim$20 K, the derived carbon masses depend very weakly on the assumed temperature \citep{weiss2005a}. \cite{papadopoulos2022} found however that the [\ion{C}{1}] lines are sub--thermally excited in the ISM of galaxies, which means that $Q_\mathrm{10}$ becomes a non--trivial function of density and temperature. Our choice of $T_\mathrm{ex}$ = 30 K and LTE only implies a 4\% higher mass compared more complex non--LTE approaches \citep{harrington2021,dunne2022,papadopoulos2022}, which is negligible compared to the uncertainty in $X_\mathrm{[CI]}$.

We obtain the CO(1--0)--based molecular gas masses following the standard equation:

\begin{equation} \label{co_gas}
    M_\mathrm{H_2}^\mathrm{CO} = \alpha_\mathrm{CO} \ L'_\mathrm{CO(1-0)} \ M_\odot,
\end{equation}

\noindent where $\alpha_\mathrm{CO}$ is the CO-to-H$_2$ conversion factor, which we take to be 1 M$_\odot$ (K km s$^{-1}$ pc$^2$)$^{-1}$ \citep[e.g.,][]{danielson2011,hodge2012,debreuck2014,chen2017,calistro-rivera2018,xue2018,riechers2020b-coldz,friascastillo2022,amvrosiadis2023}. 

Combining Eq. \ref{ci_gas} and \ref{co_gas}, we can constrain the product of the two main unknowns for each tracer (for  $Q_\mathrm{10}$ = 0.46):

\begin{equation}
    X_\mathrm{[CI]}\times\alpha_\mathrm{CO} = 2.07\times10^{-4}\frac{L'_\mathrm{[CI]}}{L'_\mathrm{CO}}.
\end{equation}

In Fig. \ref{conversion_factor} we show $X_\mathrm{[CI]}\times\alpha_\mathrm{CO}$ as a function of redshift and L$_\mathrm{IR}$, color--coded by the intensity of the FUV radiation field, $G_\mathrm{UV}$. We do not show the sources where neither CO(1--0) nor [\ion{C}{1}](1--0) are detected. The factor does not appear to be correlated with $G_\mathrm{UV}$, suggesting that it could be insensitive to local conditions of the ISM in our sample. We look for any trends in the sample with \textsc{Linmix} \citep{kelly2007}, a Bayesian linear regression fitting model which can fit data with errors along with the upper limits. Our fits return a slope of 0.02$\pm$0.02 and 0.01$\pm$0.05 for the dependence on redshift and L$_\mathrm{IR}$, respectively. The full sample has a scatter of 0.12 dex in both cases. This is consistent with no dependence on either parameter, and we can thus compute a constant conversion factor. We find a median $X_\mathrm{[CI]}\times\alpha_\mathrm{CO}$ = (5.1$\pm$1.8)$\times$10$^{-5}$ M$_\odot$ (K km s$^{-1}$ pc$^2$)$^{-1}$ and a mean (5.4$\pm$2.6)$\times$10$^{-5}$ M$_\odot$ (K km s$^{-1}$ pc$^2$)$^{-1}$ for the galaxies in our sample (not including those with both CO and [\ion{C}{1}] non--detections).

\subsubsection{Dust Continuum Emission as a Gas Mass Tracer} \label{sec:Dust}

Because the Rayleigh-Jeans (RJ) tail of the dust emission is almost always optically thin, it can trace the total dust mass, and therefore potentially the molecular gas mass, provided that the dust emissivity per unit mass and the dust-to-gas abundance ratio can be constrained. Under the assumption of a mass-weighted cold dust temperature, $T_\mathrm{dust}$ (25 K is typically considered to be representative for both local star-forming and high-redshift galaxies), and a dust emissivity index, $\beta$ (generally taken to be 1.8), the CO(1--0) luminosity and rest--frame 850 $\mu$m continuum flux have been shown to correlate for a range of galaxy populations (e.g., \citealt{scoville2016} for local SFGs, ULIRGs and high-redshift SMGs, and \citealt{kaasinen2019} for $z\sim$2 SFGs). From a single--band continuum flux measurement, the $L_\mathrm{850\mu m,rest}$ can be calculated following \cite{scoville2016,scoville2017}:

\begin{multline}
 L_\mathrm{850\mu m,rest} = 1.19\times10^{27} S_{\nu_\mathrm{obs}}(1+z)^{-(3+\beta)} \times \\ \Big(\frac{\nu_{850}}{\nu_\mathrm{obs}}\Big)^{2+\beta} \ D_\mathrm{L}^2 \ \frac{\Gamma_0(T_\mathrm{dust})}{\Gamma_{RJ}(T_\mathrm{dust},z)}  \  \mathrm{erg \ s^{-1} Hz^{-1}}, 
\end{multline}

where $\nu_\mathrm{obs}$ is the observed--frame frequency where the continuum flux density is measured and $\Gamma_\mathrm{RJ}$ is the correction for departures in the rest frame of the Planck function from Rayleigh–Jeans. The $H_2$ mass is then obtained via:

\begin{equation} \label{dust_gas}
    M_\mathrm{H_2}^\mathrm{dust} =\frac{L_\mathrm{850\mu m,rest}}{\alpha_\mathrm{850}} \ M_\odot,
\end{equation}

where $\alpha_\mathrm{850}$ is the luminosity--to--gas mass conversion factor for dust at rest--frame 850 $\mu$m. The 3 mm dust--continuum emission measurements for our sample probe rest--frame wavelengths in the range $\lambda_\mathrm{rest}\sim650-520 \mu m$, so we calculate the extrapolated $L_\mathrm{850\mu m,rest}$ and compare them against the CO and [\ion{C}{1}] line luminosities to constrain the product of their conversion factors. Firstly, we need to make an assumption for $T_\mathrm{dust}$ and $\beta$. Studies on the dust SED of SMGs find that they are characterised by warmer temperatures and steeper $\beta$ \citep[e.g.,][]{saintonge2013,dacunha2021,dunne2022,shouwen2022,AS2COSPEC_II2024} than the typically assumed $T_\mathrm{dust}$ = 25 K and $\beta=1.8$. Following \cite{AS2COSPEC_II2024}, who performed SED fitting with \textsc{CIGALE} for a subsample of the sources in this work, we set $T_\mathrm{dust}$ = 30 K and $\beta$ = 2.1 and calculate $L_\mathrm{850\mu m,rest}$. If we had chosen $T_\mathrm{dust}$ = 25 K and $\beta=1.8$ instead, the luminosities would have been 15\% higher\footnote{While the choices of $T_\mathrm{dust}$ and $\beta$ have a relatively minor effect on the derived $L_\mathrm{850\mu m,rest}$ from the observed--frame 3mm dust continuum, the impact becomes more severe if the observed wavelength does not probe the RJ tail of the dust emission. If we had used observed--frame 850 $\mu$m dust continuum observations, which are more commonly available at high redshift, the derived luminosities could have been overestimated by more than 50\%.}.

\begin{figure*}
    \centering
    \includegraphics[scale=0.4]{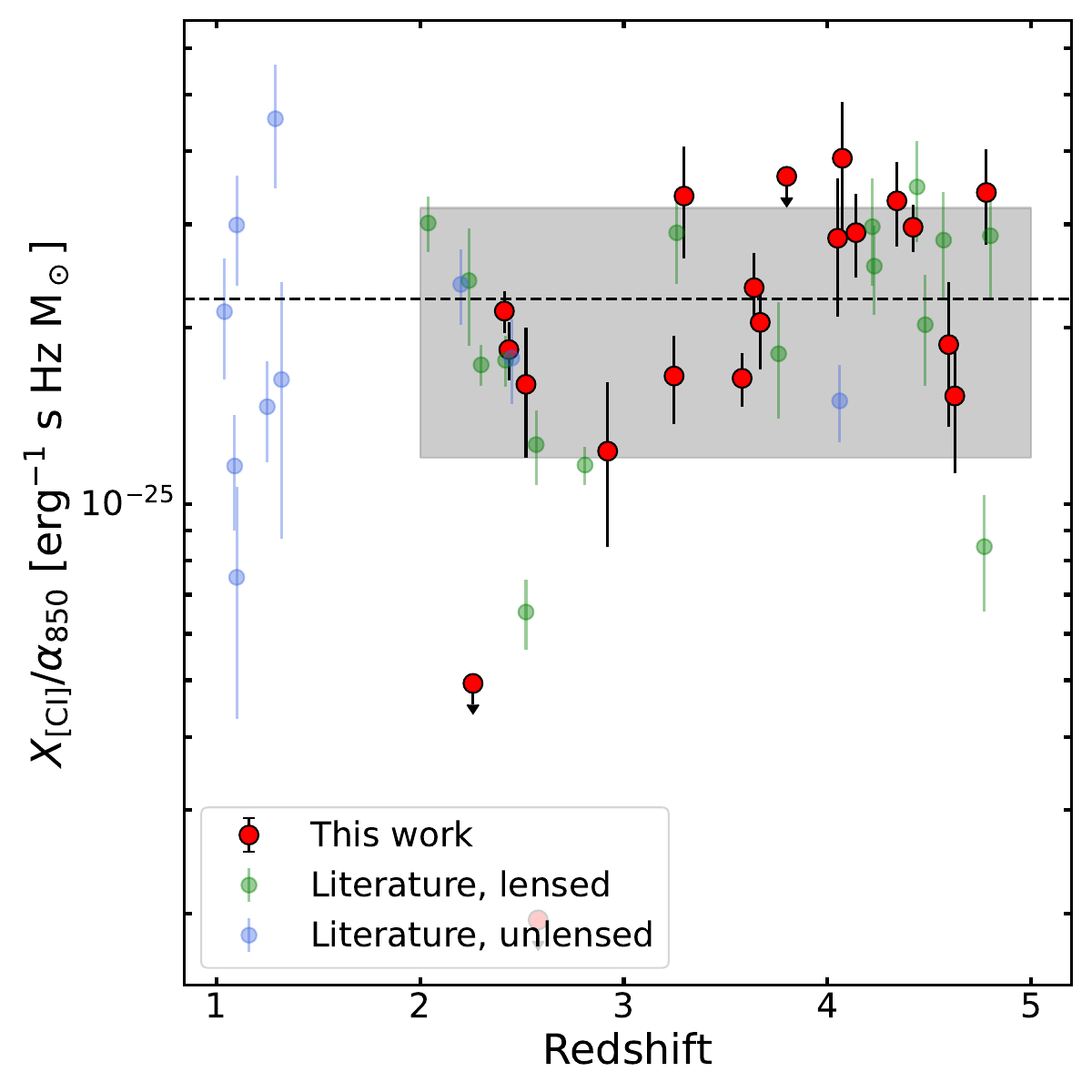}
    \includegraphics[scale=0.4]{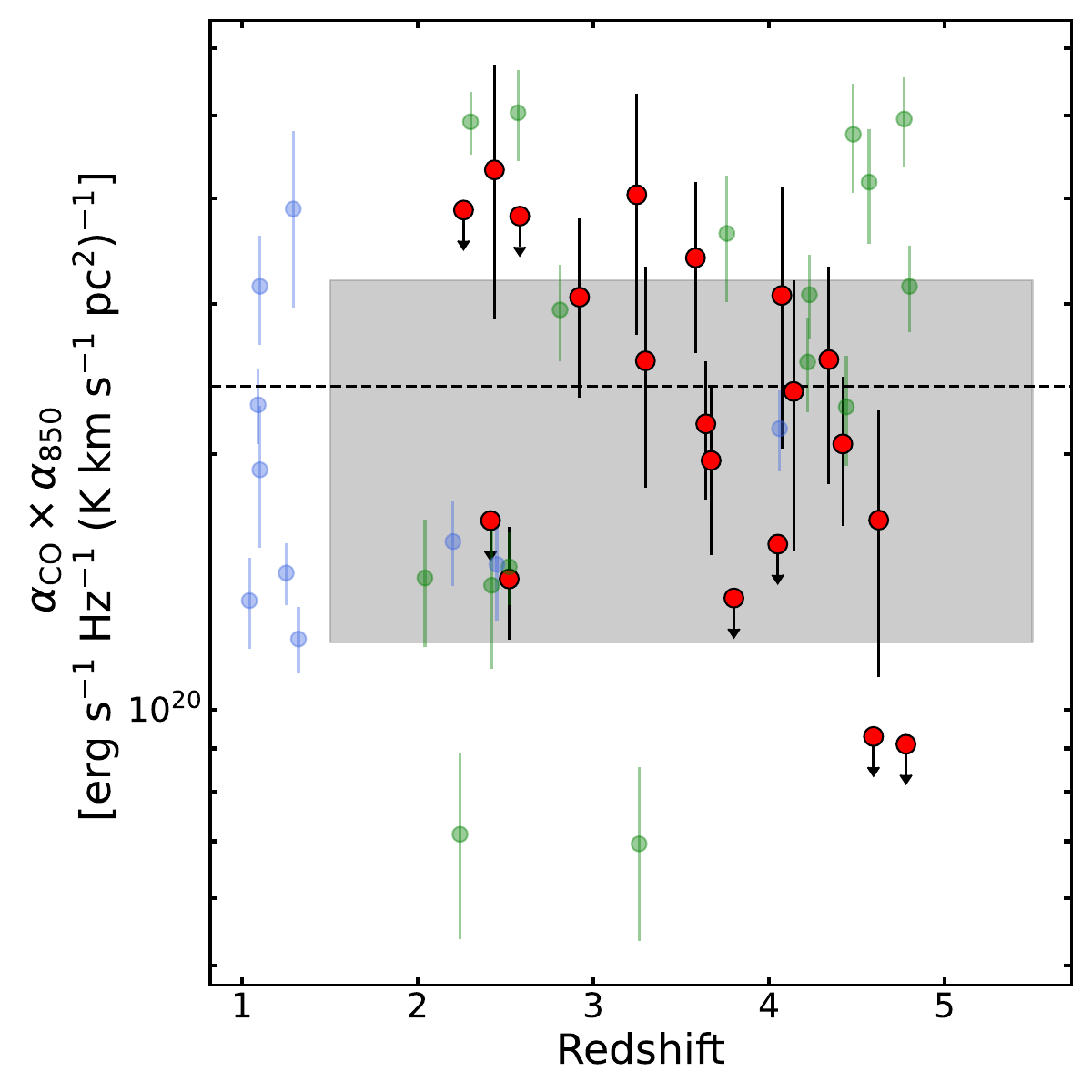}
    \caption{$X_\mathrm{[CI]}/\alpha_\mathrm{850}$ (\textit{left}) and $\alpha_\mathrm{CO}\times\alpha_\mathrm{850}$ (\textit{right}) factors as a function of redshift for our sample. The mean value for the sample in this work is shown as a black dashed line, and the gray-shaded region represents the 1$\sigma$ standard deviation around the sample mean. For comparison, we show the $z\sim1$ MS galaxies from ASPECS \citep{boogaard2020} and high--redshift SMGs \citep{walter2011,alaghband2013,bothwell2017,Canameras2018,harrington2018,nesvadba2019}. We do not find evidence for any trends with redshift or $L_\mathrm{IR}$ (not shown) for either pair.}
    \label{dust_factor}
\end{figure*}

In Fig. \ref{dust_factor} we show the distribution of $\alpha_\mathrm{CO}\times\alpha_\mathrm{850}$ and $X_\mathrm{[CI]}/\alpha_\mathrm{850}$ for our sample as a function of redshift. We fit the distributions again using \textsc{LINMIX}. 
In all cases, the fits have a slope consistent with no trend with either parameter, so we compute a mean constant conversion factor for the whole sample of:
\begin{align*}
    \alpha_\mathrm{CO}\times\alpha_\mathrm{850} = \frac{L_\mathrm{850\mu m,rest}}{L'_\mathrm{CO}} = \\ (2.4 \pm 1.0) \times 10^{20} \ \mathrm{erg} \  \mathrm{s}^{-1} \mathrm{Hz}^{-1} \ (\mathrm{K} \ \mathrm{km} \ \mathrm{s}^{-1}\ \mathrm{pc}^2)^{-1}
\end{align*}
\begin{align*}
    X_\mathrm{[CI]}/\alpha_\mathrm{850} = 2.07\times10^{-4}\frac{L'_\mathrm{[CI]}}{L_\mathrm{850\mu m,rest}} = \\ (2.2 \pm 1.0) \times 10^{-25} \ (\mathrm{erg} \ \mathrm{s}^{-1} \ \mathrm{Hz}^{-1})^{-1}.
\end{align*}


\subsection{CMB Effect on CO and [\ion{C}{1}] Emission at High Redshift} \label{sec:CMB}

\begin{figure*}[!htb]
    \centering
    \includegraphics[scale=0.28]{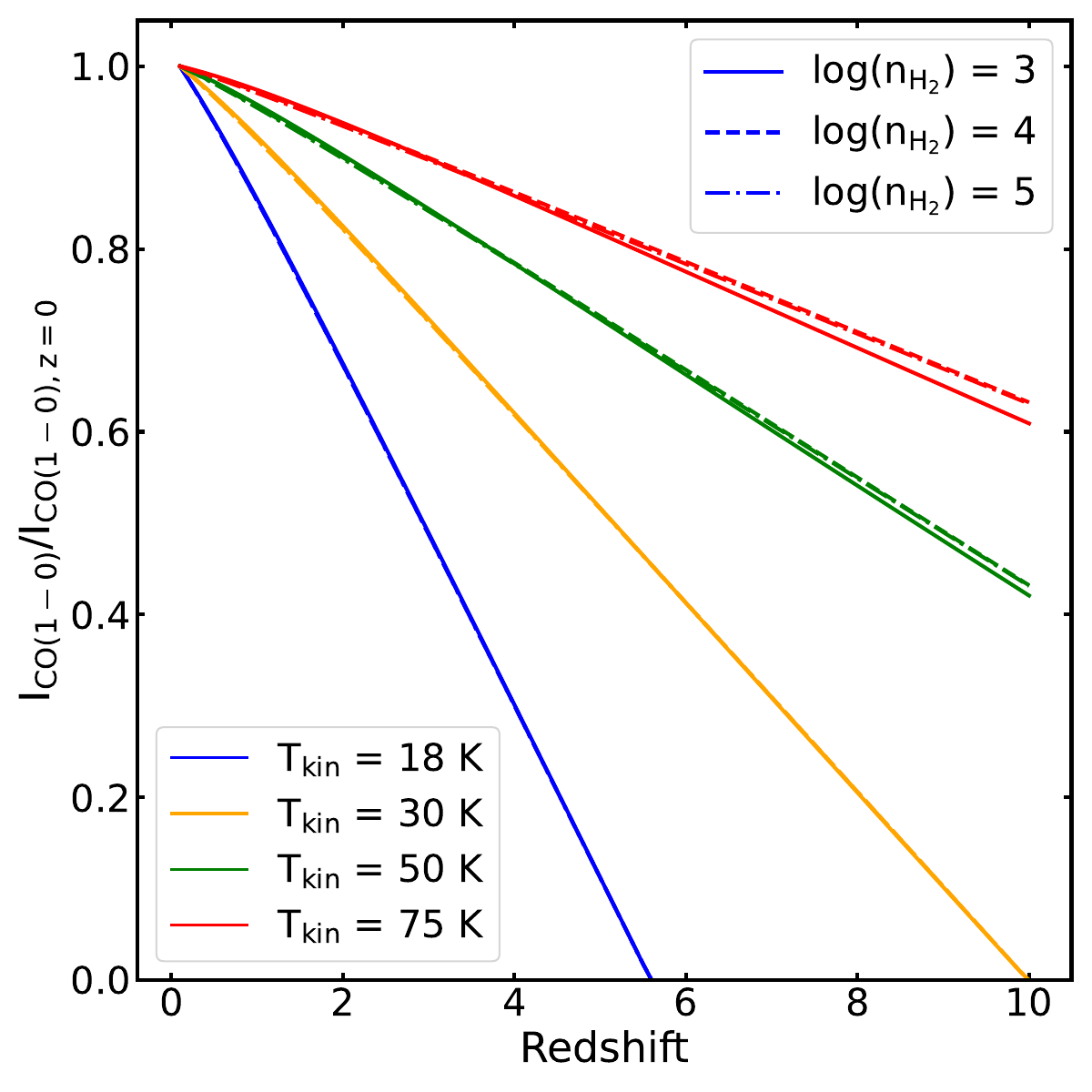}
    \includegraphics[scale=0.28]{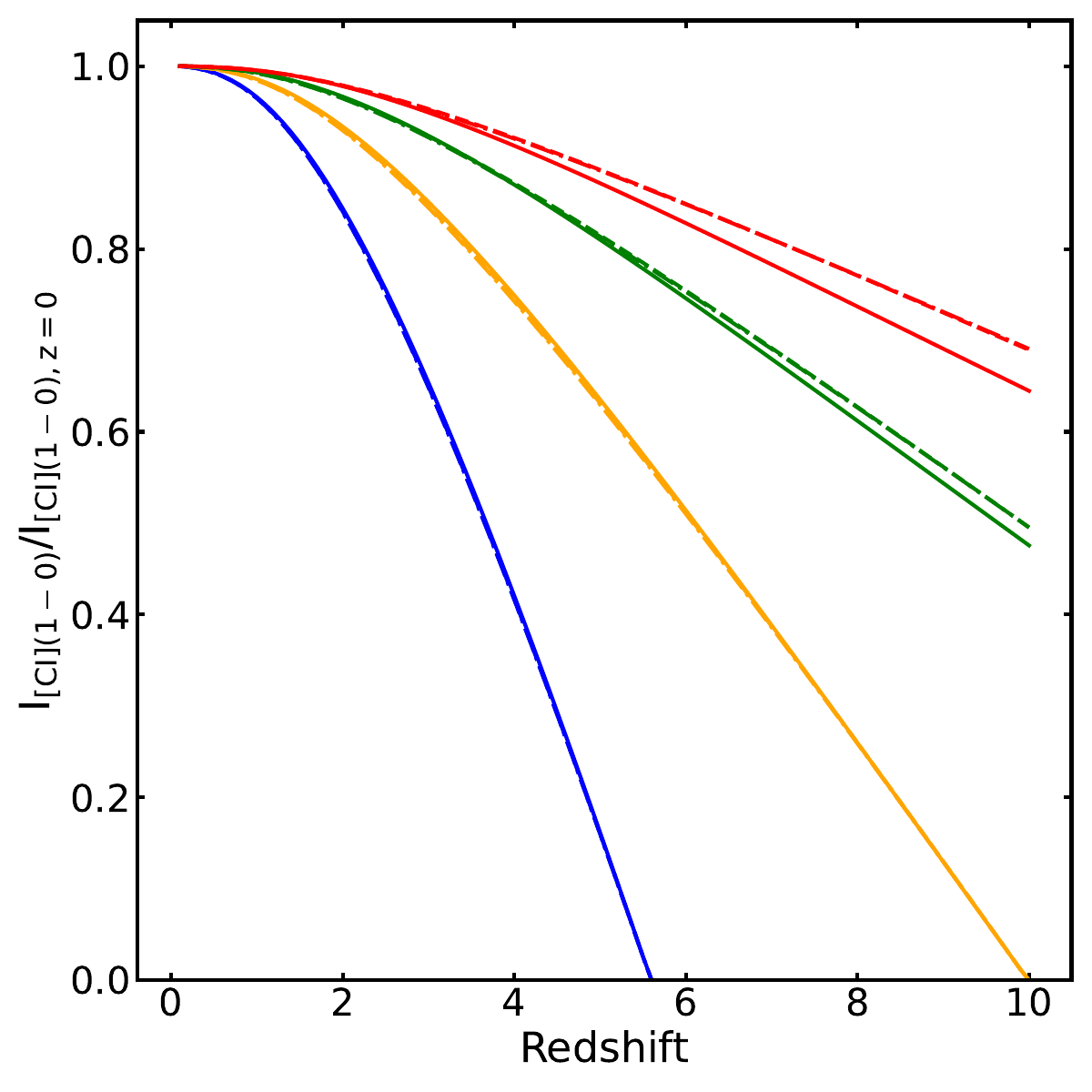}
    \includegraphics[scale=0.28]{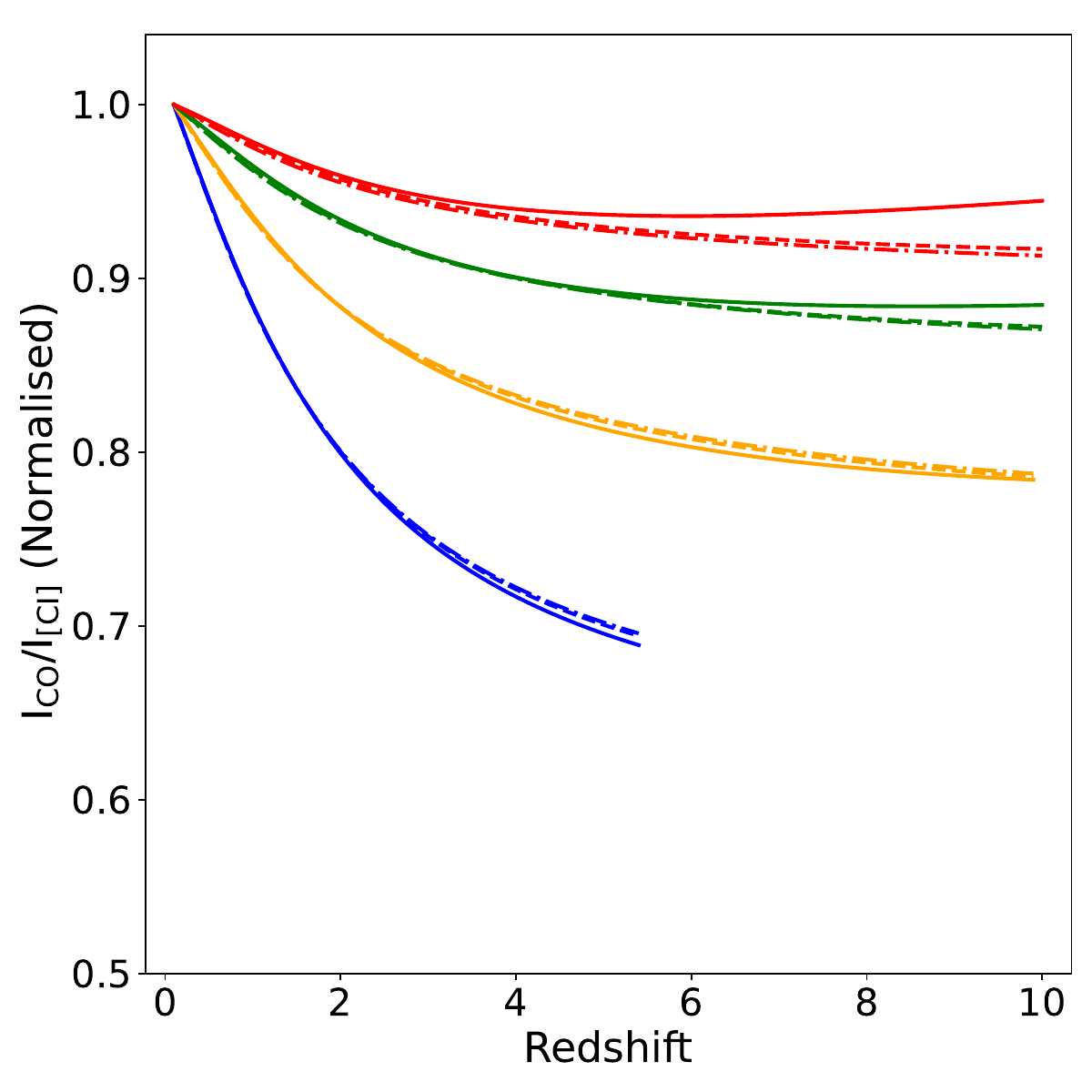}
    \caption{Recovered I$_\mathrm{CO(1-0)}$ (\textit{left}), I$_\mathrm{[CI](1-0)}$ (\textit{middle}), and their ratio (\textit{right}) as a function of redshift. For each panel, we plot the measured intensity at each redshift divided by the intrinsic intensity measured at $z=0$, where $T_\mathrm{CMB}$= 2.73 K. We show three examples of H$_2$ density, log($n$) = 3, 4 and 5, as well as a range of gas kinetic temperatures, for a column density of $N_\mathrm{CO,C}$/$d$v = 10$^{18}$.} 
    \label{ratio_z}
\end{figure*}

Our sample of SMGs spans a redshift range $\sim2-5$. Therefore, it is important to consider the effect that the CMB will have on the recoverable emission of the ground state transitions for both CO and [\ion{C}{1}], as its temperature increases \citep{dacunha2013,zhang2016}. We model how the measured CO(1--0) and [\ion{C}{1}](1--0) intensities, as well as their ratio, vary with increasing $T_\mathrm{CMB}$ for a range of gas densities and kinetic temperatures, and we show the results in Fig. \ref{ratio_z}. We use \textsc{RADEX} \citep{radex2007} for the models, a non-LTE radiative transfer code which uses the escape probability method, and we take collisional data between each molecule and H$_2$ from the LAMBDA database \citep{lambda_database}. We fix the H$_2$ ortho--to--para ratio to 3:1 and set log($N_\mathrm{CO}$/$d$v) = log($N_\mathrm{C}$/$d$v) = 18.

Our models indicate a non-negligible decrease in the recovered [\ion{C}{1}](1--0) intensity for a given gas kinetic temperature and density. The magnitude of the effect is mostly determined by the kinetic temperature of the gas, with the density having barely any impact on the resulting emission. However, the decrease in recovered line intensities for both CO(1--0) and [\ion{C}{1}](1--0) offset each other, such that there is a negligible effect on the measured luminosity ratio for the warmer gas temperatures ($T_\mathrm{kin} \geq 50 K$). For the colder temperatures, the ratio can decrease by up to $\sim$30\%. As a lower bound for the $T_\mathrm{kin}$ in our sample, we can consider the excitation temperature of the CO(1--0)-emitting layer from our PDR models; this is 18--20~K, though the real $T_\mathrm{kin}$ is likely significantly higher. 
Given the current scatter of the CO(1--0) and [\ion{C}{1}](1--0) data, it is challenging to detect a suppression of this magnitude in the ratio. We perform the same analysis for dust emission (see Appendix), and find that the CO/dust and [\ion{C}{1}]/dust flux ratios also stay approximately constant with redshift, as the effect of the CMB is of a similar order of magnitude for all tracers individually (assuming $T_\mathrm{dust}$=$T_\mathrm{kin}\sim30$ K). 

How would these effects impact the observed luminosity ratios? The warmer CMB background would act to increase the $X_\mathrm{[CI]}\times\alpha_\mathrm{CO}$ factor needed for higher--redshift sources. This is not immediately evident from
Fig. \ref{conversion_factor}, as the fit to the data is consistent with no evolution with redshift. If we separate our sources in two bins at $z<3.5$ and $z>3.5$, we find consistent mean factors for both bins (5.6$\pm$2.0$\times$10$^{-5}$ M$_\odot$ (K km s$^{-1}$ pc$^2$)$^{-1}$ versus 5.1$\pm$1.8$\times$10$^{-5}$ M$_\odot$ (K km s$^{-1}$ pc$^2$)$^{-1}$). This suggests that, at least out to $z\sim5$, the ratio of both tracers is not obviously suppressed compared to that of local ULIRGs. This could be a selection effect, as we are currently restricted to the most massive, dusty, star--forming galaxies at higher redshifts, where gas temperatures and densities, as well as metallicities, are expected to be higher than in normal, main--sequence galaxies \citep{valentino2020-co,harrington2021,Jarugula2021}. We also caution that there is a larger spread in ratios for the high--redshift sources compared to the $z=0$ ULIRGs. For the conversion factors involving dust, we find 1.5$\pm$1.0$\times$10$^{-25}$ versus 2.7$\pm$1.0$\times$10$^{-25}$ erg s$^{-1}$ Hz$^{-1}$ (K km s$^{-1}$ pc$^{2}$)$^{-1}$ for $\alpha_\mathrm{CO}\times\alpha_\mathrm{850}$, and 1.8$\pm$0.8$\times$10$^{20}$ versus 3.1$\pm$0.8$\times$10$^{20}$ K km s$^{-1}$ pc$^{2}$ (erg s$^{-1}$ Hz$^{-1}$)$^{-1}$ for $X_\mathrm{[CI]}/\alpha_\mathrm{850}$ for the lower and higher redshift bins, respectively -- again consistent with no evolution with redshift.

\section{Discussion}\label{sec:Discussion}

The observations of CO(1--0) and [\ion{C}{1}](1--0) presented here for a sample of dusty, star--forming galaxies at $z\sim2-5$ obviate the need for excitation corrections, allowing us to directly confirm that the relation found in the local Universe remains in place for the ground state transition of both species out to $z\sim5$. This suggests that [\ion{C}{1}] coexists with CO in the cold gas reservoirs of high--redshift galaxies and is therefore a good tracer of the total molecular gas mass. However, as we do not yet have a calibration of [\ion{C}{1}] as a tracer of molecular gas mass fully independent of CO; we can only constrain the product $X_\mathrm{[CI]}\times\alpha_\mathrm{CO}$ and relate it to the choice of $\alpha_\mathrm{CO}$. We find a median $X_\mathrm{[CI]}\times\alpha_\mathrm{CO}$ = (5.1$\pm$1.8)$\times$10$^{-5}$ M$_\odot$ (K km s$^{-1}$ pc$^2$)$^{-1}$ and a mean of (5.4$\pm$2.6)$\times$10$^{-5}$ M$_\odot$ (K km s$^{-1}$ pc$^2$)$^{-1}$ (without including three sources that are undetected in both lines). This mean product is consistent with the value found for the SPT lensed galaxies by \cite{gururajan2023} of (6.3$\pm$0.7)$\times$10$^{-5}$ M$_\odot$ (K km s$^{-1}$ pc$^2$)$^{-1}$ and the value obtained in the study by \cite{dunne2022} of (3.98$\pm$0.19)$\times$10$^{-5}$ M$_\odot$ (K km s$^{-1}$ pc$^2$)$^{-1}$, which included a compilation of published CO(1--0) and [\ion{C}{1}](1--0) observations available at that time.

\begin{figure*}[!htb]
    \centering
    \includegraphics[width=0.48\linewidth]{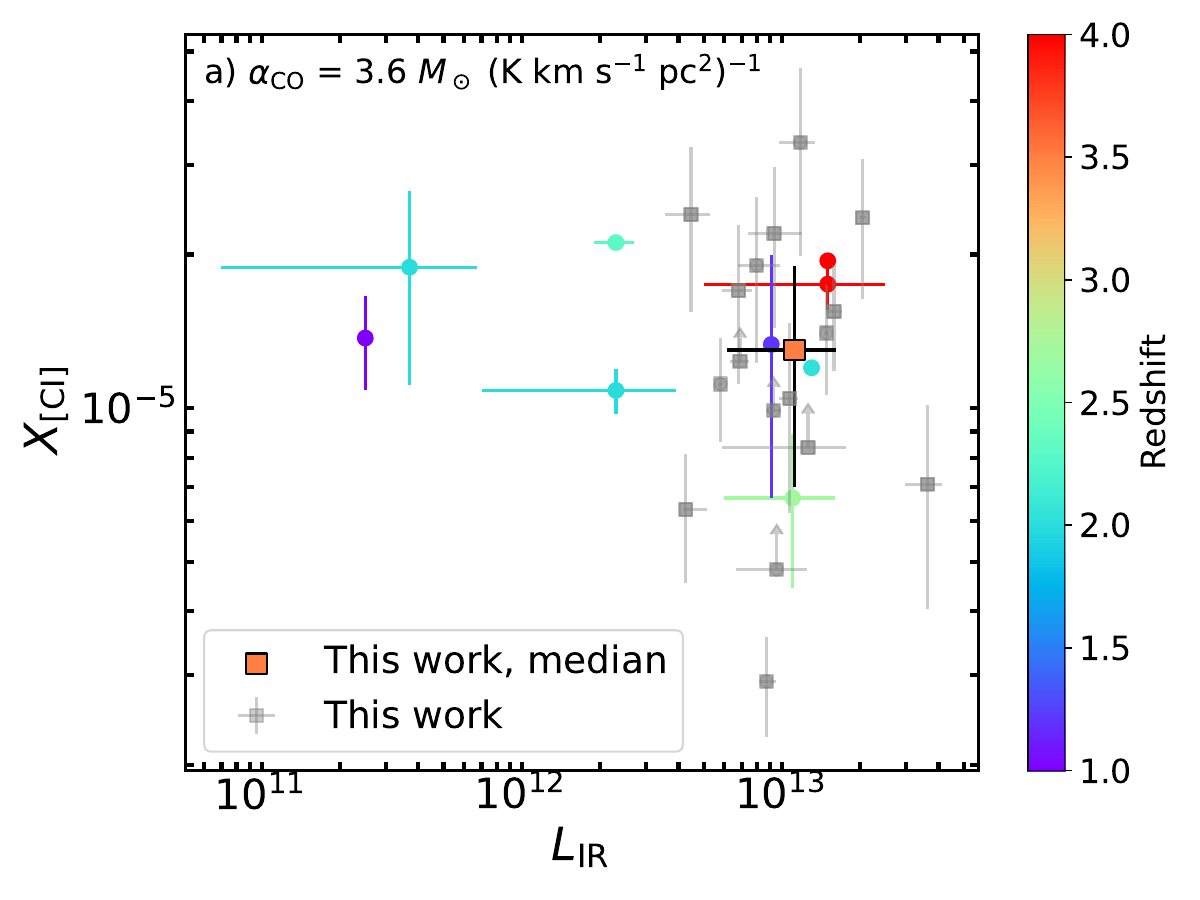}
    \includegraphics[width=0.48\linewidth]{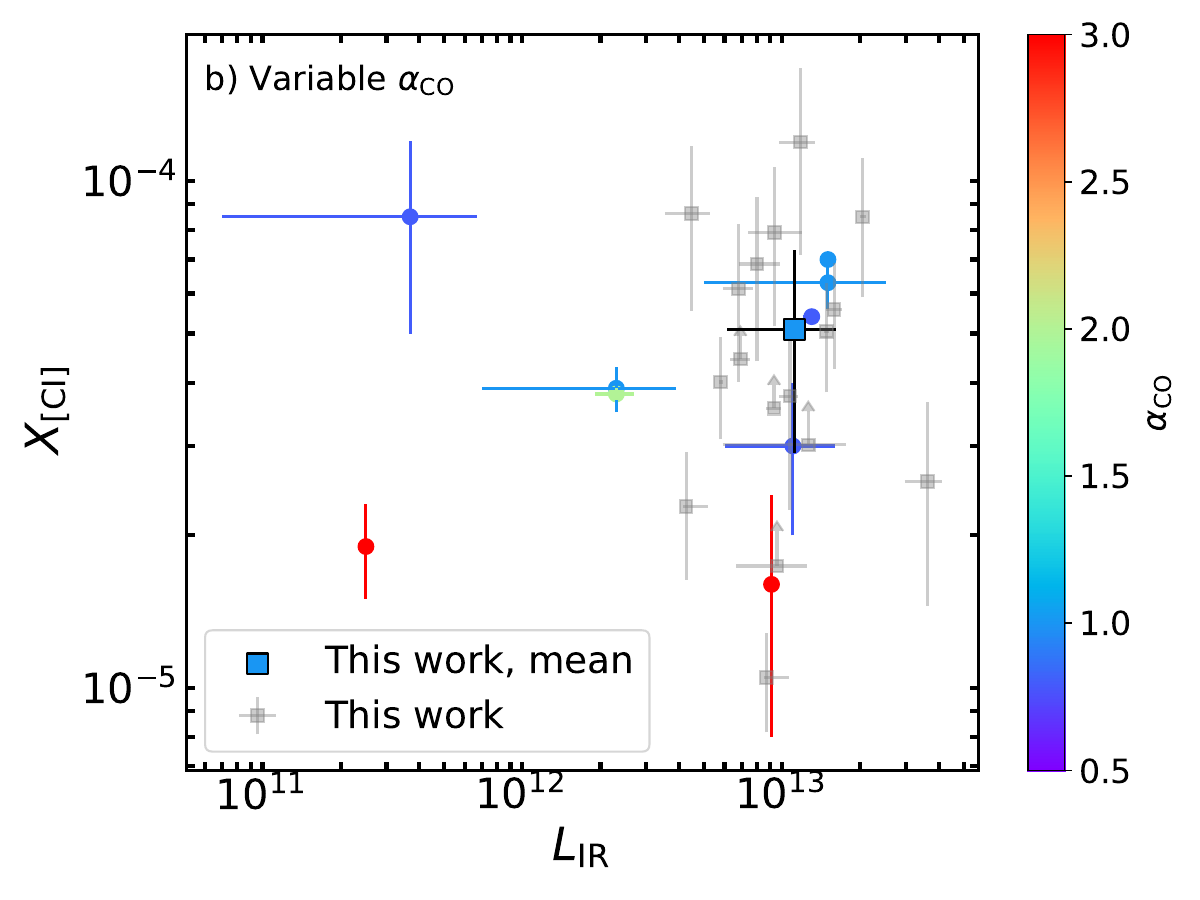}
    \caption{\textit{Left}: Values of $X_\mathrm{[CI]}$ reported in the literature as a function of the average $L_\mathrm{IR}$ of the corresponding sample. We show the average of our sample and plot the individual points for each galaxy in gray. We show sources from \cite{danielson2011,walter2011,alaghband2013,bothwell2017,valentino2018,dannerbauer2019,boogaard2020,nesvadba2019,gururajan2023}, colour--coded by the average redshift of the sample. The values have been corrected for the same $\alpha_\mathrm{CO}=3.6$ \citep{dunne2022}. \textit{Right}: Same as the panel on the left, but the points are colour--coded by the value of $\alpha_\mathrm{CO}$ used to derive each carbon abundance. The introduction of a bi--modal $\alpha_\mathrm{CO}$ naturally leads to an equal bi--modality in $X_\mathrm{[CI]}$.}
    \label{final_fig}
\end{figure*}

 Assuming a Milky--Way value of $\alpha_\mathrm{CO}$=3.6, we derive a median $X_\mathrm{[CI]}=(1.4\pm0.5)\times10^{-5}$, which would be in agreement with abundances derived for local \citep{jiao2019} and $z\sim1$ main--sequence galaxies \citep{valentino2018,boogaard2020}, as well as the sample from \cite{dunne2022} ($1.6^{+0.5}_{-0.4}\times10^{-5}$). 
 On the other hand, if we adopt $\alpha_\mathrm{CO}\sim$1, as is common for high--redshift SMGs and local ULIRGs, we obtain a mean carbon abundance of $X_\mathrm{[CI]}=(5.1\pm1.8)\times10^{-5}$, higher than the commonly adopted value of 3$\times$10$^{-5}$ \citep{weiss2003,papadopoulos2004,bothwell2017} or the values found for main sequence galaxies. Higher carbon abundances have been derived for local ULIRGs \citep{jiao2017,jiao2019} and high--redshift SMGs \citep{walter2011,alaghband2013,gururajan2023}, although the latter have commonly targeted lensed sources and relied on $J_\mathrm{up}>2$ CO observations. These higher values can be caused by regions with higher cosmic ray ionization rates or stronger FUV radiation field (Fig.~\ref{pdr}), which would dissociate CO into [\ion{C}{1}] \citep{bisbas2021}. Higher carbon abundances could also be explained by galaxies having higher metallicities -- we return to this point below.  
 
 Finally, assuming $X_\mathrm{[CI]}=5.1\times10^{-5}$ and $\alpha_\mathrm{CO}=1$, we derive consistent mean values for the dust conversion factor $\alpha_\mathrm{850}=(2.2\pm1.0)\times10^{20}$ erg s$^{-1}$ Hz$^{-1}$ M$_\odot^{-1}$ and $(2.4\pm1.0)\times10^{20}$ erg s$^{-1}$ Hz$^{-1}$ M$_\odot^{-1}$, respectively. This is a factor of two lower than the mean value of $(4.4\pm1.1)\times10^{20}$ erg s$^{-1}$ Hz$^{-1}$ M$_\odot^{-1}$ derived in \cite{scoville2016} (correcting their value for $\alpha_\mathrm{CO}=1$ and removing the Helium contribution, as we are only dealing with $M_\mathrm{H_2}$).

Previous studies reported a discrepancy between the $H_2$ masses derived from CO and [\ion{C}{1}] observations \citep[e.g.,][]{bothwell2017,valentino2018,montoya2023}. This can be explained by the fact that these studies used a bi--modal $\alpha_\mathrm{CO}$ while keeping the carbon abundance constant for all galaxy populations (X$_\mathrm{[CI]}$ = 3$\times$10$^{-5}$), which naturally results in a mismatch in the derived H$_2$ masses. We show this effect in Fig.~\ref{final_fig}, where we have compiled the carbon abundances reported in previous studies. When corrected to the same $\alpha_\mathrm{CO}$, the values of X$_\mathrm{[CI]}$ show a fairly linear distribution with $L_\mathrm{IR}$ and redshift, as previously noted by \citet{dunne2022}. However, if we adopt different values for $\alpha_\mathrm{CO}$ for different galaxy populations, studies  that adopt a ULIRG--like value generally find higher carbon abundances than those that use a Milky--Way value instead. 

\begin{figure}[!htb]
    \hspace{-1cm}
    \includegraphics[scale=0.45]{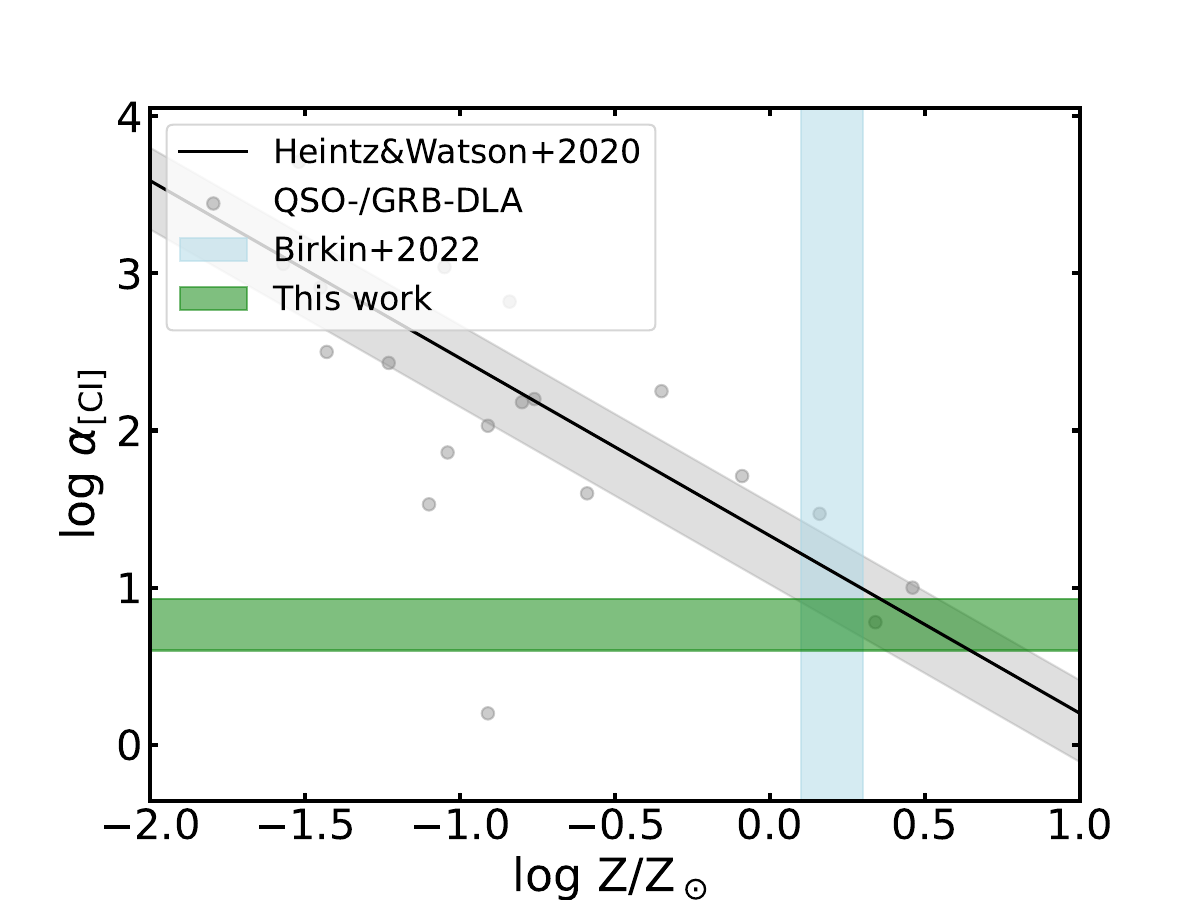}
    \caption{[CI] conversion factor $\alpha_\mathrm{[CI]}$ as a function of metallicity. The dark green band marks the value derived using $X_\mathrm{[CI]}$=(5.1$\pm$1.8)$\times$10$^{-5}$ for the sample of SMGs presented in this work. The light green band marks the SMG metallicities derived from [\ion{N}{2}]/H$\alpha$ by \cite{birkin2022PhDT.........6B}. The black line represents the best-fit linear relation log $\alpha_\mathrm{[CI]}$ = -1/13 $\times$ log $Z/Z_\odot$ + 1.33 from \cite{heintz_watson2020}. The solar/super--solar metallicities derived support the use of a higher (lower) $X_\mathrm{[CI]}$ ($\alpha_\mathrm{CO}$) in SMGs, rather than a universal value for different galaxy populations.}
    \label{metallicity_fig}
\end{figure}

Although \citet{dunne2022} argued for near--universal values of X$_\mathrm{[CI]}$ = 1.6$\times$10$^{-5}$ and $\alpha_\mathrm{CO}$ = 3.6 M$_\odot$ (K km s$^{-1}$ pc$^2$)$^{-1}$, multiple high--resolution studies of gas kinematics in SMGs have noted the difficulty of reconciling such a high  $\alpha_\mathrm{CO}$ with the derived stellar and dynamical masses, consistently finding that $\alpha_\mathrm{CO}\sim1$ (and therefore X$_\mathrm{[CI]}\sim$ 5$\times$10$^{-5}$)  is preferred for this population \citep[e.g.,][see also \cite{magdis2011} for a similar conclusion on the $z=4.05$ SMG GN20 using dust constraints]{danielson2011,hodge2012,debreuck2014,chen2017,calistro-rivera2018,xue2018,birkin2020,riechers2020b-coldz,friascastillo2022,amvrosiadis2023}. Following \cite{heintz_watson2020}, the values of  $X_\mathrm{[CI]}$ derived for our sources assuming $\alpha_\mathrm{CO}=1$ would imply solar or super--solar metallicities (Fig.~\ref{metallicity_fig}). 

This is in line with SMGs being dust enriched and is consistent with recent results from rest-frame optical spectroscopy with VLT's KMOS \citep[which includes SMGs from the AS2UDS and AS2COSMOS surveys]{birkin2022PhDT.........6B} and JWST's NIRSpec \citep{Birkin2023}, although some SMGs might have sub-solar metallicities (e.g., \citealt{Rigopoulou2018, Rybak2023}). 
This re-affirms our choice of $\alpha_\mathrm{CO}$ and derived $X_\mathrm{[CI]}$. Assuming $X_\mathrm{[CI]}\sim$5$\times$10$^{-5}$, we derive gas masses $M_\mathrm{H_2,[CI]}$ in the range 5--21$\times10^{10} M_\odot$, consistent with $M_\mathrm{H_2,CO}$ = 4--13$\times10^{10} M_\odot$ using $\alpha_\mathrm{CO}=1$. Therefore, it is possible to obtain gas mass estimates from either tracer that agree within the uncertainties, so long as consistent assumptions are made when converting luminosities into gas masses. Finally, we note that, due to the relative brightness of [\ion{C}{1}] compared to CO(1--0), [\ion{C}{1}] will thus provide more precise gas mass estimates than CO(1--0) for the same integration time (though subject to the same systematic uncertainties). 

The constant conversion factor that we have derived relies on the linear correlation between $L'_\mathrm{CO}$ and $L'_\mathrm{[CI]}$ observed out to $z\sim 5$, as well as the lack of evolution of the ratio of both quantities with either redshift or $L_\mathrm{IR}$. We have shown that the warmer CMB at higher redshifts has a moderate impact on the ratio, which can be suppressed by up to $30\%$ for the lowest kinetic gas temperatures (Fig.~\ref{ratio_z}). In Fig.~\ref{lci_lum} (right) we show the impact that this could have on the measured ratio at high redshift. We have taken the mean value of the line luminosity ratio $L'_\mathrm{[CI]}$/$L'_\mathrm{CO}$ for local ULIRGs (0.21) as the intrinsic value that an SMG would have at $z=0$, and applied the correction for the effect of the CMB for different $T_\mathrm{kin}$. The effect on the ratio is too small to be differentiated from the current scatter of the data, which is likely driven by the uncertainties in the measurements as well as the internal conditions of individual galaxies \citep{gururajan2023}. We caution that this only applies to the ratio of both tracers, and that the CMB effect can be significant on individual lines, which in turns affects the absolute value of the molecular gas masses that we obtain at high redshift. In order to determine the absolute decrease of the line intensities due to the CMB of individual lines, it will be necessary to obtain more complete SLEDs to perform complex, non--LTE modelling of the gas excitation conditions \citep{harrington2021} and extend the sample to galaxies of lower mass and metallicities.

\section{Conclusions} \label{sec:Conclusions}

We have presented ALMA Band 3 and 4 observations targeting [\ion{C}{1}](1--0) in a sample of twelve $z\sim2-5$ unlensed, massive, gas--rich galaxies with CO(1--0) observations from a new VLA survey of molecular gas in massive star-forming galaxies at high redshift \citep{friascastillo2023}. Combining these data with eight archival detections in the CO(1--0) sample, we have compiled a total of 20 galaxies with observations of the ground state transitions of both species. This dataset allows us to directly compare CO(1--0) and [\ion{C}{1}](1--0) as tracers of the cold molecular gas reservoirs in galaxies at high redshift. The main conclusions are as follows:

\begin{itemize}
    \item  We detect [\ion{C}{1}](1--0) line emission in 10 out of 12 galaxies observed. The full sample has [\ion{C}{1}](1--0) line luminosities in the range 1.2--3.1$\times$10$^{10}$ K km s$^{-1}$ pc$^2$. The [\ion{C}{1}](1--0)/CO(1--0) line luminosity ratio is consistent with the one found at lower redshifts for ULIRGs and main--sequence galaxies, and shows no deviation out to $z\sim5$. 

    \item Combining the mid--$J_\mathrm{up}$ CO and [\ion{C}{1}] line luminosities with infrared luminosities, we constrain the conditions of the ISM of our sample using PDR models. We find a median density log$(n \ [$cm$^{-3}])=4.7\pm0.2$, and a median UV radiation field log$(G_{\mathrm{UV}} \ [G_0]) = 3.2\pm0.2$. There is a significant overlap in the parameter space occupied by our sources and local ULIRGs, $z\sim1$ main sequence galaxies and published high--redshift SMGs.

    \item We compare our measurements of $L'_\mathrm{CO(1-0)}$, $L'_\mathrm{[CI](1-0)}$ and 3mm dust continuum to provide a cross--calibration for their conversion factors, $\alpha_\mathrm{CO}$, $X_\mathrm{[CI]}$, and $\alpha_\mathrm{850}$, respectively. The conversion factors do not evolve with redshift or $L_\mathrm{IR}$. We find that the product $X_\mathrm{[CI]}\times\alpha_\mathrm{CO}$ is remarkably similar for local ULIRGs, main sequence galaxies and high--redshift SMGs, and this affects the derived $X_\mathrm{[CI]}$ values when different values of $\alpha_\mathrm{CO}$ are assumed for various galaxy populations. Taking $\alpha_\mathrm{CO}=1$ for our sample of SMGs, we derive a median $X_\mathrm{[CI]}=(5.1\pm1.8)\times10^{-5}$. This results in gas masses $M_\mathrm{H_2,[CI]}$ in the range 5--21$\times10^{10} M_\odot$. Likewise, these choices of $\alpha_\mathrm{CO}$ and $X_\mathrm{[CI]}$ result in values of $\alpha_\mathrm{850}=(2.2\pm1.0)\times10^{20}$ erg s$^{-1}$ Hz$^{-1}$ M$_\odot^{-1}$ and $(2.4\pm1.0)\times10^{20}$ erg s$^{-1}$ Hz$^{-1}$ M$_\odot^{-1}$, respectively. These values differ from the canonically assumed $X_\mathrm{[CI]}=3\times10^{-5}$ and $\alpha_\mathrm{850}=4.4\times10^{20}$ erg s$^{-1}$ Hz$^{-1}$ M$_\odot^{-1}$, but are supported by predictions for the solar/super--solar metallicities expected for our sources.

    \item We model the effect of the warmer CMB at high redshift on the measured intensities of CO(1--0) and [\ion{C}{1}](1--0). The absolute effect on both individual lines can be quite severe, but the relative effect between the lines is much smaller, such that the ratio between the lines decreases by up to $\sim30\%$ for the coldest gas kinetic temperatures expected. A comparison of the $L'_\mathrm{CO(1-0)}/$$L'_\mathrm{[CI](1-0)}$ ratio between local ULIRGs and our sources shows both populations have similar values, suggesting that the ratio is not obviously suppressed out to $z\sim5$.
\end{itemize}

We caveat that current constraints on the conversion factors at high redshift are limited to typically very active sources. To improve the calibration of $X_\mathrm{[CI]}$, $\alpha_\mathrm{CO}$ and $\alpha_\mathrm{850}$ at high redshift, as well as the constrains on the effects of the CMB, it is necessary to obtain samples of galaxies spanning a larger range of masses and metallicities. Additionally, well--sampled CO and [\ion{C}{1}] SLEDs will facilitate a more accurate modeling of properties of their cold ISM. The galaxies presented in this work constitute the largest sample of un-lensed galaxies with CO(1--0) and [\ion{C}{1}(1--0)] observations, which will be crucial for understanding the physical conditions of cold ISM at high redshift.

\acknowledgments
This paper makes use of the following ALMA data: ADS/JAO.ALMA\#2021.1.01342.S. ALMA is a partnership of ESO (representing its member states), NSF (USA) and NINS (Japan), together with NRC (Canada), MOST and ASIAA (Taiwan), and KASI (Republic of Korea), in cooperation with the Republic of Chile. The Joint ALMA Observatory is operated by ESO, AUI/NRAO and NAOJ. This paper makes use of data from IRAM/NOEMA program S19CV (PI: Chapman). M. R. is supported by the NWO Veni project ``\textit{Under the lens}'' (VI.Veni.202.225). J.H. acknowledges support from the ERC Consolidator Grant 101088676 (VOYAJ) and the VIDI research programme with project number 639.042.611, which is (partly) financed by the Netherlands Organisation for Scientific Research (NWO).
 J.B. has received funding from the European Research Council (ERC) under the European Union’s Horizon 2020 research and innovation programme MOPPEX 833460. Part of this work was supported by the German \emph{Deut\-sche For\-schungs\-ge\-mein\-schaft, DFG\/} project number Ts~17/2--1. T.T. gratefully acknowledges the Collaborative Research Center 1601 (SFB 1601 subproject A6) funded by the Deutsche Forschungsgemeinschaft (DFG, German Research Foundation) – 500700252. C.C.C and C.-L.L. acknowledge support from the National Science and Technology Council of Taiwan (NSTC  111-2112M-001-045-MY3), as well as Academia Sinica through the Career Development Award (AS-CDA-112-M02). IRS and AMS acknowledge support from STFC (ST/T000244/1 and ST/X001075/1). E.F.-J.A. acknowledge support from UNAM-PAPIIT project IA102023, and from CONAHCyT Ciencia de Frontera project ID:  CF-2023-I-506. HD acknowledges financial support from the Agencia Estatal de Investigaci\'on del Ministerio de Ciencia e Innovación (AEI-MCINN) under grant (La evoluci\'on de los c\'umulos de galaxias desde el amanecer hasta el mediod\'ia c\'osmico) with reference (PID2019-105776GB-I00/DOI:10.13039/501100011033)  and the Ministerio de Ciencia, Innovaci\'on y Universidades (MCIU/AEI) under grant (Construcci\'on de c\'umulos de galaxias en formaci\'on a trav\'es de la formaci\'on estelar oscurecida por el polvo) and the European Regional Development Fund (ERDF) with reference (PID2022-143243NB-I00/DOI:10.13039/501100011033). This work was supported by NAOJ ALMA Scientific Research Grant Code 2021-19A (H.S.B.A). The National Radio Astronomy Observatory is a facility of the National Science Foundation operated under cooperative agreement by Associated Universities, Inc. The authors acknowledge assistance from Allegro, the European ALMA Regional Center node in the Netherlands. 

\appendix
\addcontentsline{toc}{section}{Appendices}

\subsection*{CMB Effect on Dust Emission}\label{appendix}
\begin{figure}[!htb]
    \centering
    \includegraphics[scale=0.35]{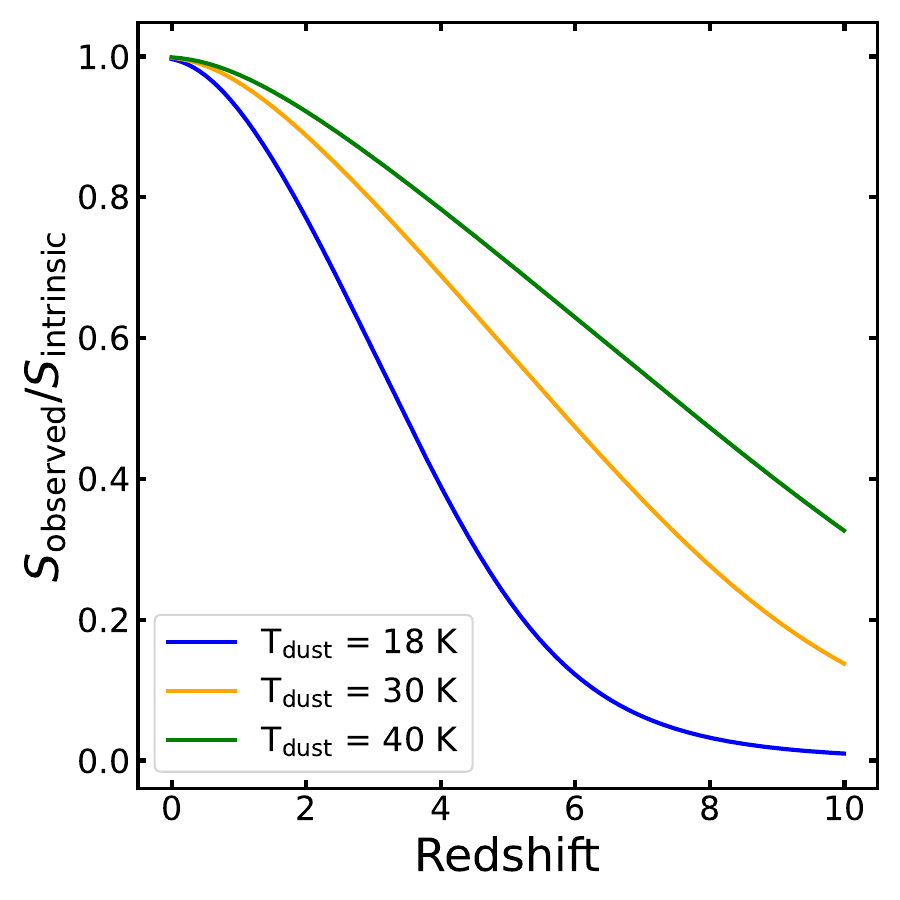}
    \includegraphics[scale=0.35]{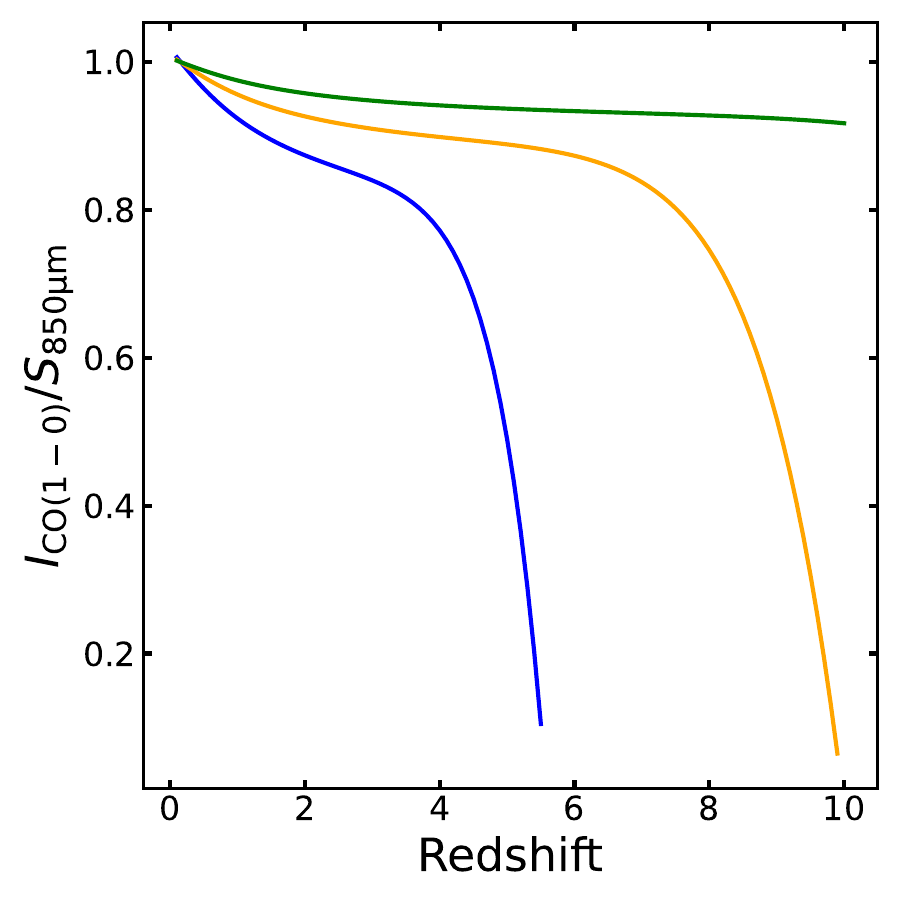}
    \includegraphics[scale=0.35]{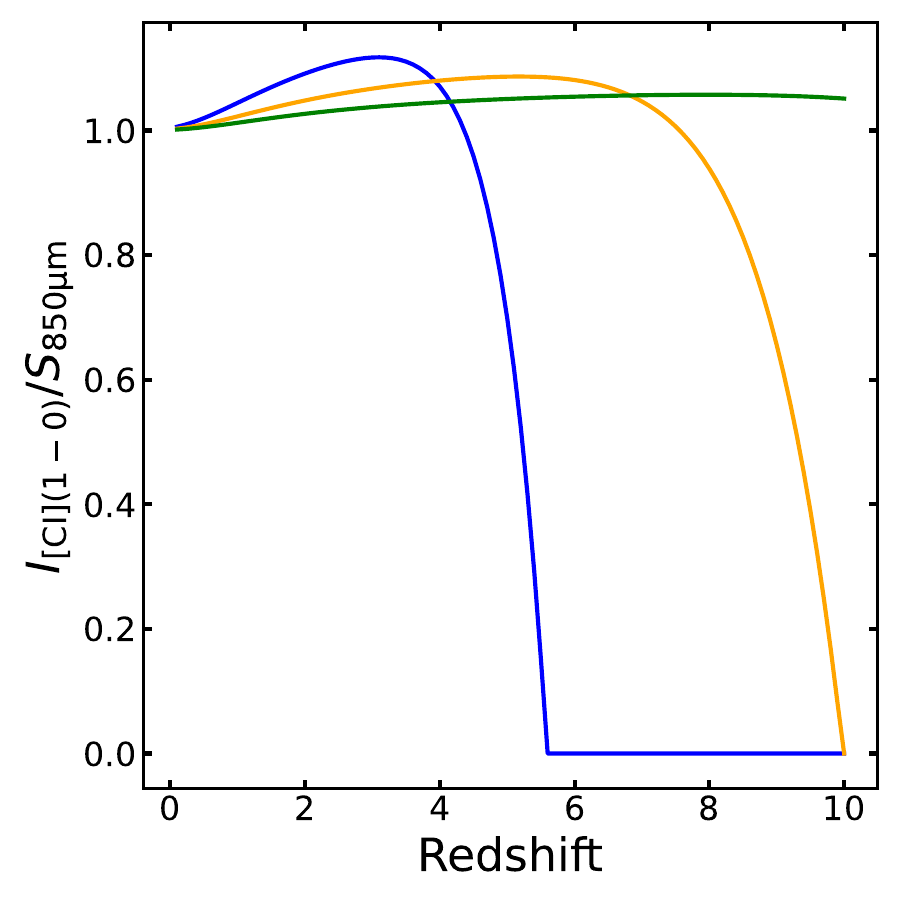}
    \caption{Effect of the warmer CMB on the recovered dust emission at 100 GHz as a function of redshift. We assume three different $T_\mathrm{dust}$ at $z=0$, and plot the ratio between the intrinsic flux emitted by that galaxy at $z=0$ and what can be recovered as the warmer CMB increases in temperature at higher redshift.} 
    \label{ratio_dust}
\end{figure}

\bibliography{main}{}
\bibliographystyle{aasjournal}

\end{document}